\documentclass[12pt]{article}

\ifx\pdfoutput\undefined
\usepackage[dvips,bookmarks]{hyperref}
\else
\usepackage{hyperref}
\fi
\hypersetup{colorlinks=false,bookmarksopen,bookmarksnumbered,citecolor=blue,
   pdfstartview=FitH}

\usepackage{latexsym}
\usepackage{amssymb,amsfonts,amsmath}
\usepackage{graphicx} 
\usepackage{indentfirst}
\usepackage{bbm}

\parskip=\medskipamount

\newcommand {\cD}{{\cal D}}
\newcommand {\cE}{{\cal E}}

\newcommand {\cG}{{\cal G}}

\newcommand {\cJ}{{\cal J}}

\newcommand {\cN}{{\cal N}}

\newcommand {\cS}{{\cal S}}
\newcommand {\cT}{{\cal T}}

\newcommand {\cV}{{\cal V}}
\newcommand {\cW}{{\cal W}}
\newcommand {\cX}{{\cal X}}
\newcommand {\cY}{{\cal Y}}

\newcommand{\bG}{{\bf G}}
\newcommand{\bH}{{\bf H}}

\newcommand{\bV}{{\bf V}}
\newcommand{\bW}{{\bf W}}

\def\a{\alpha}
\def\b{\beta}
\def\c{\chi}
\def\d{\delta}

\def\f{\phi}
\def\g{\gamma}

\def\k{\kappa}
\def\l{\lambda}

\def\o{\omega}

\def\q{\theta}

\def\s{\sigma}

\def\x{\xi}
\def\z{\zeta}
\def\D{\Delta}
\def\F{\Phi}
\def\J{\Psi}
\def\L{\Lambda}
\def\O{\Omega}

\def\U{\Upsilon}

\def \bi{\bibitem}

\def\ri{{\rm i}}

\newcommand{\ad}{{\dot{\alpha}}}                           
\newcommand{\bd}{{\dot{\beta}}}                            
\newcommand{\ve}{\varepsilon}                            
\newcommand{\cDB}{{\bar\cD}}                            

\newcommand{\pa}{\partial}                           
\newcommand{\hf}{\frac12}

\newcommand{\vf}{\varphi}

\newcommand{\be}{\begin{equation}}
\newcommand{\ee}{\end{equation}}
\newcommand{\bea}{\begin{eqnarray}}
\newcommand{\eea}{\end{eqnarray}}
\newcommand{\non}{\nonumber}
\newcommand{\ba}{\begin{array}}
\newcommand{\ea}{\end{array}}

\newcommand{\1}{\underline{1}}
\newcommand{\2}{\underline{2}}

\newcommand{\bm}[1]{\mbox{\boldmath$#1$}}

\def\double #1{#1{\hbox{\kern-2pt $#1$}}}

\newcommand{\gd}{{\dot\g}}

\newcommand{\CD}{{\nabla}}

\newcommand{\bsubeq}{\begin{subequations}}
\newcommand{\esubeq}{\end{subequations}}

\newcommand{\ul}{\underline}
\newcommand{\eps}{{\epsilon}}

\newcommand{\dalpha}{{\dot{\alpha}}}
\newcommand{\dbeta}{{\dot{\beta}}}

\newcommand{\btheta}{{\bar\theta}}
\newcommand{\N}{{\mathcal N}}
\newcommand{\eol}{\notag \\}
\newcommand{\rd}{\mathrm d}
\newcommand{\qW}{{\mathbf W}}
\newcommand{\qG}{{\mathbf G}}
\newcommand{\qH}{{\mathbf H}}
\newcommand{\qOmega}{{\mathbf \Omega}}
\newcommand{\FF}{\nabla}
\newcommand{\bFF}{\bar \nabla}
\newcommand{\qomega}{{\bm\omega}}

\begin{document}

\begin{titlepage}
\begin{flushright}
16 November 2010\\
\end{flushright}
\vspace{5mm}

\begin{center}
{\Large \bf $\bm{\cN = 2}$ supergravity and supercurrents}
\\ 
\end{center}

\begin{center}

{\bf Daniel Butter and  Sergei M. Kuzenko }

\footnotesize{
{\it School of Physics M013, The University of Western Australia\\
35 Stirling Highway, Crawley W.A. 6009, Australia}}  ~\\
\texttt{dbutter,\,kuzenko@cyllene.uwa.edu.au}\\
\vspace{2mm}

\end{center}
\vspace{5mm}

\begin{abstract}
\baselineskip=14pt
We address the problem of classifying all $\cN=2$ supercurrent multiplets in four space-time dimensions.  
For this purpose we consider the minimal formulation of $\cN=2$ Poincar\'e supergravity with a tensor compensator, 
and derive its linearized action in terms of three $\cN=2$  off-shell multiplets: 
an unconstrained  scalar superfield, a vector multiplet, and a tensor multiplet.
Such an action was ruled out  to exist
in the past. Using the action constructed, one can derive other models for linearized $\cN=2$ supergravity 
by applying $\cN=2$ superfield duality transformations.
The action depends parametrically on a constant real isotriplet 
$g^{ij}=g^{ji}\neq 0$ which originates as  an expectation value of the tensor compensator. Upon reduction to $\cN=1$ superfields, we show that the model describes two dually equivalent formulations for the massless multiplet
$(1,3/2)\oplus (3/2, 2)$ depending on a choice of $g^{ij}$. In the case $g^{\1\1}=g^{\2\2}=0$, 
the action describes (i) new minimal $\cN=1$ supergravity; and (ii) the Fradkin-Vasiliev-de Wit-van Holten
gravitino multiplet. In the case $g^{\1\2}=0$, on the other hand, the action describes (i) old minimal $\cN=1$ supergravity; and (ii) the Ogievetsky-Sokatchev gravitino multiplet. 
\end{abstract}

\vfill
\end{titlepage}

\newpage
\renewcommand{\thefootnote}{\arabic{footnote}}
\setcounter{footnote}{0}

\tableofcontents


\numberwithin{equation}{section}


\newpage
\section{Introduction}
\setcounter{footnote}{0}
The supercurrent \cite{FZ} is a supermultiplet which contains the conserved
energy-momentum tensor and the conserved supersymmetry current, along with
some other components including the $R$-current (which is not always conserved).
Its fundamental significance is due to the fact that this multiplet embraces all
the conserved currents associated with the super-Poincar\'e symmetry.
In complete analogy with the energy-momentum tensor, which is the source of gravity,
the supercurrent is the source of supergravity \cite{OS,FZ2}.

The structure of supercurrent multiplets in $\cN=1$ supersymmetric theories is
fully understood. The supercurrent can be consistently derived by varying the matter
action in a curved superspace with respect  to the supergravity prepotentials,
and then restricting to the flat superspace background. Since there exist 
several off-shell formulations for $\cN=1$ supergravity (specifically, the old
minimal \cite{old}, the new minimal  \cite{new} and the non-minimal \cite{non-min}
formulations), they lead to different supercurrent multiplets (see textbooks
\cite{GGRS,BK} for pedagogical reviews) of which the Ferrara-Zumino
multipet \cite{FZ} corresponds to old minimal supergravity.
The technique of deriving the supercurrent from off-shell supergravity, 
which was described in detail in \cite{GGRS,BK}, can be streamlined and 
re-formulated as a superfield Noether procedure \cite{Osborn,MSW}. 
Recently, there has been much interest in various aspects of the supercurrents 
emerging in gauge theories with Fayet-Iliopoulos terms and nonlinear sigma-models 
with non-exact K\"ahler forms \cite{KS,DT,K-FI,KS2,K-var,Butter,K-Noet}
inspired by the work  of Komargodski and Seiberg \cite{KS,KS2}.

Unlike the case of simple supersymmetry, the structure of supercurrents in
$\cN=2$ supersymmetric theories is much less studied. In the early
papers \cite{Sohnius,HST,Stelle}, supercurrents were studied on a model-dependent basis.
At that time it was practically impossible to derive supercurrents starting from
supergravity, since only  the discovery of harmonic superspace \cite{GIKOS} made
possible the construction of fully-fledged prepotential formulations for $\cN=2$
supergravity \cite{Galperin:1987em,Galperin:1987ek} (see \cite{GIOS} for a review.)
The supergravity origin of several $\cN=2$ supercurrent multiplets was revealed
in \cite{KT}.

The oldest $\cN=2$ supercurrent multiplet was introduced by Sohnius \cite{Sohnius}
by considering the hypermultiplet. The supercurrent is described by a real scalar
superfield $\cJ$. The associated trace supermultiplet is an isotriplet,
$\cT^{ij} =\cT^{ji}$, constrained by
\be
D^{(i}_\a \,\cT^{jk)} = {\bar D}^{(i}_\ad \,\cT^{jk)} = 0~,
 \qquad  ( \cT^{ij} )^* =\cT_{ij} =\ve_{ik}\ve_{jl}\cT^{kl}~.
\label{trace}
\ee
Neither the supercurrent nor the multiplet of anomalies have a central charge.
The constraints (\ref{trace}) are characteristic of the $\cN=2$ linear
multiplet \cite{BS,SSW}. The supercurrent conservation equation
is\footnote{The conservation equation (\ref{sccl}) can be rewritten in a
different  form ${\bar D}^{ij} \hat{\cJ} =D^{ij} \hat{\cJ} = 4 \hat{\cT}^{ij}$,
where $\hat{\cT}^{ij}$ is a real isotriplet obeying the constraint
$D^{(i}_\a \, \hat{\cT}^{jk)} = {\bar D}^{(i}_\ad \, \hat{\cT}^{jk)} = 0$.
This form is obtained  from (\ref{sccl}) in two steps. First, one
represents $\cT^{ij} =- {\rm i} \,D^{ij} \J +{\rm i} \,{\bar D}^{ij} {\bar \J} $, 
for some chiral superfield  $\J$. Secondly, one defines the modified supercurrent
$ \hat{ \cJ } :=  \cJ +8 ( \J +\overline{ \J } )$. It obeys the conservation
equation postulated, with $\hat{\cT}^{ij} := D^{ij} \J + {\bar D}^{ij} {\bar \J} $.}
\bea
\frac{1}{4} {\bar D}^{ij} \cJ = {\rm i}\, \cT^{ij} = -\frac{1}{4}D^{ij} \cJ~,
\label{sccl}
\eea 
where $D^{ij} := D^{\a  i} D^{j}_\a =D^{ji}$ and 
${\bar D}^{ij} := {\bar D}^{(i}_\ad {\bar D}^{j) \ad }={\bar D}^{ji}$. 

It is worth giving a couple of examples of $\cN=2$ supersymmetric theories in 
which the supercurrent has the type just described. First of all, we consider
the $\cN=2$  Maxwell action \cite{GSW}
\begin{align}
S_{\rm V} =\frac{1}{2} \int \rd^4x\, \rd^4\theta\, W^2~, \qquad {\bar D}_i^\ad W=0~,
\end{align}
where $W$ is the chiral field strength of an Abelian vector multiplet. It obeys
the Bianchi identity
\begin{align}\label{eq_BianchiW}
D^{ij} W = \bar D^{ij} \bar W~.
\end{align}
The field strength can be constructed in terms of Mezincescu's prepotential
\cite{Mezincescu}, $V_{ij}$, which is  a real unconstrained $SU(2)$ triplet:
\be
W ={\bar D}^4 D^{ij} V_{ij}~, \qquad V_{ij} = V_{ji}~, \qquad
(V_{ij} )^* = V^{ij} =\ve^{ik}\ve^{jl} V_{kl}~.
\label{Mezincescu}
\ee
The supercurrent for this model \cite{HST} is
\bea
\cJ = W \bar W ~, \qquad \cT^{ij}=0~.
\eea

A more interesting example is given by the low-energy effective action \cite{rsg}
\bea
S =  \int \rd^4x\, \rd^4\theta\, F( W^I) +  \int \rd^4x\, \rd^4{\bar \q} \, {\bar F}( {\bar W}^I )~,
\label{holomor}
\eea
with $F$ a holomorphic function of $n$ variables. For this theory, 
the supercurrent and the trace supermultiplet are 
\begin{subequations}
\bea
\cJ &=& {\bar W}^I F_I (W) + W^I {\bar F}_I (W) +  \Big\{ W^I F_I( W) -2F( W) 
+{\rm c.c.} \Big\}~, \label{holomor1}\\
\cT^{ij}&=& \frac{\rm i}{4} D^{ij} \Big\{ W^I F_I (W) -2 F(W)  \Big\} +{\rm c.c.}
\label{holomor2}
\eea
\end{subequations}
In a somewhat different form, this supercurrent was derived in \cite{MSW}.

Now, consider the $\cN=2$ vector multiplet with a Fayet-Iliopoulos term
\bea
S_{\rm V+FI} =\frac{1}{2} \int \rd^4x\, \rd^4\theta\, W^2 
- \int \rd^4x\, \rd^4\theta\, {\rm d}^4\bar \q  \,\x^{ij} V_{ij}~, \qquad 
\x^{ij}={\rm const}~.
\eea
The corresponding equation of motion is $D^{ij} W = \x^{ij}$.
One can see that the only gauge invariant candidate for the supercurrent is again 
$\cJ =W \bar W$. However, the supercurrent conservation equation becomes
\bea
{\bar D}^{ij} \cJ = \x^{ij} W~.
\eea
It is obvious that this conservation equation differs from that in eq. (\ref{sccl}).
The supergravity origin of this difference is very simple. As shown in \cite{KT}, 
the conservation equation (\ref{sccl}) occurs in those $\cN=2$ supersymmetric 
theories which couple only to the minimal multiplet of $\cN=2$ supergravity \cite{BS}
(i.e., the  Weyl multiplet \cite{deWvHVP,BdeRdeW} coupled to an Abelian vector multiplet,
the latter being the first supergravity compensator), and do not couple to a second 
supergravity compensator. In the case under consideration, the Fayet-Iliopoulos term
directly couples to the second compensator in the off-shell formulation for $\cN=2$
Poincar\'e supergravity proposed in \cite{deWPV}. Its second compensator is 
the improved $\cN=2$ tensor multiplet \cite{deWPV, LR} which is a natural generalization
of the improved $\cN=1$ tensor multiplet \cite{deWR}.

The above example provides enough rational for investigating general $\cN=2$ supercurrent
multiplets. In the case of $\cN=1$ supersymmetry, the variant supercurrents are naturally
associated with linearized off-shell formulations for $\cN=1$ supergravity. Given such a
formulation, the supercurrent conservation equation can be obtained by coupling the supergravity 
prepotentials to external sources and then demanding the resulting action to be invariant under 
the linearized supergravity gauge transformations. Since the linearized off-shell $\cN=1$
supergravity actions have been classified \cite{GKP}, all consistent supercurrents can be
generated. This has been carried out in \cite{K-var}. We wish to extend the $\cN=1$
construction to the $\cN=2$ case. Our approach to addressing this problem consists in
constructing a linearized superfield action for $\cN=2$ supergravity formulation proposed
in \cite{deWPV} at the component level. Other linearized supergravity actions can be
obtained from the one constructed below by superfield duality transformations.
We should mention that some models for linearized $\cN=2$ supergravity appeared in 
the early 1980s \cite{RT,GS82} shortly before Ref. \cite{deWPV} appeared. 
We will comment on these later on.

The supergravity formulation of \cite{deWPV} was originally derived in components.
Its reformulation in superspace is necessary for our goals. There exist two fully-fledged 
manifestly supersymmetric settings to describe general $\cN=2$ supergravity-matter systems:
(i) the harmonic superspace approach  \cite{Galperin:1987em,Galperin:1987ek,GIO}
(see \cite{GIOS} for a review); and (ii) the projective superspace approach 
\cite{KLRT-M1,KLRT-M2,K-dual08}. We will use both of them in the present paper.
 
In the projective superspace approach \cite{KLRT-M1,KLRT-M2,K-dual08}, 
the action for pure $\cN=2$ Poincar\'e supergravity with tensor compensator \cite{deWPV}
consists of two terms
\begin{subequations}\label{eq_N2sugra}
\bea
S_{\rm SUGRA} & = & S_{\rm minimal} +S_{\rm tensor}~, \label{1.11a}\\
S_{\rm minimal}&=&
- \frac{1}{2\k^2} \int \rd^4 x \,{\rm d}^4\q \, \cE \, {\cal W}^2 
=- \frac{1}{4 \k^2} \int \rd^4 x \,{\rm d}^4\q \, \cE \, {\cal W}^2 +{\rm c.c.}~, 
\label{1.11b} \\ 
S_{\rm tensor}&=& 
\frac{1}{2\pi \k^2} \oint_C  v^i \rd v_i
\int \rd^4 x \,{\rm d}^4\q {\rm d}^4{\bar \q}
\,\frac{E}{S^{(2)} \breve{S}^{(2)}}\,
{\cG}^{(2)} 
\ln \frac{{\cG}^{(2)}}{{\rm i}\breve{ \U}^{(1)}  \U^{(1)}}~,
\label{1.11c}
\eea
\end{subequations}
with $\k$ the gravitational coupling constant.
Here the first term, $S_{\rm minimal}$, corresponds to the minimal supergravity multiplet \cite{BS}. 
It describes the coupling of the Weyl supergravity multiplet  
\cite{deWvHVP,BdeRdeW} to an Abelian vector multiplet (with a wrong sign for the kinetic term).  
The Weyl multiplet is described using Howe's superspace geometry \cite{Howe} (see also \cite{Muller})
elaborated in detail in \cite{KLRT-M2}. The vector multiplet is described by 
a covariantly chiral field strength $\cW$  and its conjugate $\bar \cW$,
\bea
\cDB^\ad_i \cW= 0~, \qquad
\Big( \frac{1}{ 4}\cD^{\a(i}\cD_\a^{j)}+S^{ij}\Big) \cW&=&
\Big( \frac{1}{ 4}\cDB_\ad{}^{(i}\cDB^{j) \ad}+\bar{S}^{ij}\Big)\bar{\cW} ~,
\label{1.12}
\eea
where $S^{ij} $ and ${\bar S}^{ij} $  are special dimension-1 components of the torsion, 
see \cite{KLRT-M2} for more details. Finally, the superfield $\cE$  in (\ref{1.11b}) 
denotes  the chiral density, see \cite{Muller,KT-M} for its definition.

The second term in the supergravity action, $S_{\rm tensor}$, 
describes the coupling of the Weyl multiplet to an improved tensor multiplet
(with wrong sign for the kinetic term).  
The action involves a closed contour integral over auxiliary isotwistor variables 
$v^i \in  {\mathbb C}^2 \setminus \{0\}$. The tensor compensator is described by its field 
strength $\cG^{ij}$, which appears as $\cG^{(2)}$ and obeys covariant constraints, namely
\bea
\cG^{(2)}(v):= \cG^{ij} \,v_i v_j~, \qquad (\cG^{ij})^* = \cG_{ij} ~, 
\qquad \cD^{(i}_\a \cG^{jk)} =  {\bar \cD}^{(i}_\ad \cG^{jk)} = 0~.
\label{1.14}
\eea
The superfields $S^{(2)}$ and $\breve{S}^{(2)}$ in (\ref{1.11c}) are defined as
\bea
S^{(2)}(v):= S^{ij} \,v_i v_j ~\qquad \breve{S}^{(2)}(v):= {\bar S}^{ij}  \,v_i v_j ~.
\eea
As usual, $E= {\rm Ber}(E_M{}^A)$ denotes the full  superspace density,  
with $E_M{}^A$ being the super-vielbein. Finally, $\U^{(1)} (v)$ denotes a weight-1 covariant arctic 
hypermultiplet and $\breve{ \U}^{(1)} (v)$ its smile-conjugate (see \cite{KLRT-M2} for the definition 
of smile conjugation). The superfields ${ \U}^{(1)} $  and $\breve{ \U}^{(1)} $ are purely gauge degrees
of freedom, as proved in \cite{KT-M}. This property is analogous 
to that characteristic of the $\cN=1$ improved tensor multiplet \cite{deWR}, that is: 
the corresponding action in Minkowski superpsace
\be
S \propto \int \rd^4 x \,{\rm d}^4\q \,G \ln \big({G}/{\f \bar \f}\big)~, \qquad {\bar D}^2 G =D^2 G =0
\ee
does not depend on the chiral superfield $\f$ or its conjugate.
The supergravity action $S_{\rm SUGRA} $ is invariant under arbitrary supergravity gauge and super-Weyl transformations (see \cite{KLRT-M2} for more details).

Our goal is to linearize the action $S_{\rm SUGRA} $ around Minkowski superspace which is 
an exact solution of  $\cN=2$ Poincar\'e supergravity. The vector and the tensor compensators 
must be characterized by 
{\it non-vanishing} constant background values
related to each other by
\be
\cW \bar \cW= \sqrt{\frac{1}{2} \cG^{ij}\cG_{ij}}~, \qquad \cW \neq 0~,
\label{1.16}
\ee
which is the equation of motion for the gravitational superfield.
This equation has a natural counterpart at the component  level  \cite{deWPV}.
In what follows, we set $\k=1$.

This paper is organized as follows. In section 2 we elaborate on the supergravity 
origin of various $\cN=2$ supercurrent multiplets building on the results obtained in \cite{KT}.
In section 3 we derive the linearized action for $\cN=2$ supergravity. Its reduction to $\cN=1$ superfields
is carried out in section 4. Some implications of our results are discussed in section 5.
Appendix A provides a brief review of the $\cN=1$ superfield formulation \cite{LR} 
for the improved $\cN=2$ tensor multiplet.
Appendix B contains the  technical details of the 
$\cN=1$ reduction.

\section{The supergravity origin of supercurrents}
\setcounter{footnote}{0}

As shown in \cite{KT}, in  harmonic superspace  
the minimal supergravity multiplet \cite{BS} can be  described by 
two prepotentials, $H(z,u)$ and $V_5^{++} (z,u)$.
The gravitational superfield $H$ describes the  Weyl multiplet 
 \cite{deWvHVP,BdeRdeW}.
It is a real unconstrained superfield with the Fourier
expansion\footnote{The harmonics $(u_i^-, u_i^+) \in SU(2)$
obey $u^{i +} u_i^- = 1$, $u_i^+ = \eps_{ij} u^{j+}$, and
$(u^{i+})^* = u_i^-$.}
\bea
H(z,u) &=& \bH(z) + \sum_{n=1}^{\infty} 
H^{(i_1 \cdots i_n j_1\cdots j_n)} (z)
u^+_{i_1} \cdots u^+_{i_n} 
u^-_{j_1} \cdots u^-_{j_n} =\breve{H}(z,u)~.
\eea
The second prepotential $V_5^{++}$ is a real analytic
superfield of $U(1)$ charge $+2$,
\be
D^+_\a V_5^{++} ={\bar D}^+_\ad V_5^{++}=0~, \qquad 
\breve{V}_5^{++} =V_5^{++}~,
\ee
where $D_\alpha^\pm = u_i^\pm D_\alpha^i$ and $\bar D_\dalpha^\pm = u_i^\pm \bar D_\dalpha^i$.
It describes an Abelian vector multiplet which gauges the central charge that can be interpreted as 
the derivative in an extra bosonic coordinate $x^5$. The central charge gauge field 
can be represented by a Fourier series\footnote{In what follows, we use several conventions 
conventions for products of spinor derivatives, specifically: $D^4 :=\frac{1}{48} D^{ij}D_{ij}$, 
$(D^+)^4:= \frac{1}{16} (D^+)^2({\bar D}^+)^2$ and $(D^-)^4:= \frac{1}{16} (D^-)^2({\bar D}^-)^2$.
} 
\bea
V_5^{++}(z,u) &=&(D^{+})^4 U^{--}(z,u)~, \non \\
U^{--}(z,u) &=& 
\bV^{ij} (z)\,u^-_i u^-_j +
\sum_{n=2}^{\infty} 
U^{(i_1 \cdots i_{n-1} j_1\cdots j_{n+1})}(z) \,
u^+_{i_1} \cdots u^+_{i_{n-1}} 
u^-_{j_1} \cdots u^-_{j_{n+1}}  ~,~~~
\eea
with $\bV^{ij}$ Mezincescu's prepotential (compare with eq. (\ref{Mezincescu})).

The minimal supergravity multiplet  is characterized by three types of gauge symmetries.
Here we present their linearized form only; see \cite{KT} for the complete discussion.
First of all, we have the so-called  pre-gauge invariance 
\bea
\d H & = & \frac{1}{4} (D^+)^2 \O^{--} + \frac{1}{4}
({\bar D}^+)^2 \breve{\O}^{--} ~, \non \\
\d V_5^{++} &=& -\bar w (D^+)^4 \O^{--}
- w (D^+)^4 \breve{ \O}^{--} ~,
\label{pre-gauge}
\eea
with the parameter $\O^{--}(z,u)$ an unconstrained complex harmonic superfield, 
\bea
\O^{--}(z,u) &=& \sum_{n=1}^{\infty} 
\O^{(i_1 \cdots i_{n-1} j_1\cdots j_{n+1})} (z)\,
u^+_{i_1} \cdots u^+_{i_{n-1}} 
u^-_{j_1} \cdots u^-_{j_{n+1}}~.
\eea
The constant $w$ is the vacuum value of the first compensator $\cW$ in (\ref{1.11b}).
In \cite{KT}, the super-Weyl and local $U(1)$ gauge 
\be w = i
\ee
 was chosen. We use the same gauge in the present section, but we will keep
$w$  arbitrary in section 3.

The second gauge freedom corresponds to linearized general coordinate
transformations\footnote{In  the central basis, the harmonic derivative 
$D^{++}$  is defined by its action 
on the harmonics as $D^{++} u_i^- = u_i^+, D^{++} u_i^+ = 0$, and similarly for $D^{--}$.}
\bea
\d H= -D^{++} l^{--}~, \qquad \d V_5^{++} =0~,
\label{gf2}
\eea
with the gauge parameter $l^{--}$ being an unconstrained real harmonic superfield,
\bea
l^{--}(z,u) &=& \sum_{n=1}^{\infty} 
l^{(i_1 \cdots i_{n-1} j_1\cdots j_{n+1})} (z)\,
u^+_{i_1} \cdots u^+_{i_{n-1}} 
u^-_{j_1} \cdots u^-_{j_{n+1}} =\breve{l}^{--}(z,u)~.
\eea
Finally, we have the vector multiplet gauge freedom
\bea
\d H=0~, \qquad \d V_5^{++} = -D^{++} \l~,  \qquad D^+_\a \l ={\bar D}^+_\ad \l=0~,
\label{gf3}
\eea
with the gauge parameter $ \l(z,u) =\breve{\l}(z,u)$ being real analytic but otherwise arbitrary.

The gauge freedom (\ref{gf2}) can be used to choose the gauge condition\footnote{It was shown 
for the first time  in \cite{HST,RT} 
that  the linearized $\cN=2$ Weyl multiplet can be described by a real unconstrained 
prepotential $\bH$. The origin of such a  prepotential in the harmonic superspace approach to $\cN=2$ 
supergravity was revealed in \cite{Siegel-curved} at the linearized level, and in \cite{KT} at the fully nonlinear  
level.}
\be
D^{++} H = 0\quad \Longleftrightarrow \quad H(z,u) = \bH (z)~.
\label{firstgc}
\ee
The surviving gauge freedom (which we will call the ``supergravity gauge
transformation'') consists of those  combined transformations 
(\ref{pre-gauge}) and (\ref{gf2}) which 
preserve the above gauge condition, that is
\be
 \d \bH(z) = \frac{1}{12} D_{ij} \O^{ij}(z) + 
\frac{1}{12} {\bar D}_{ij} {\bar \O}^{ij}(z) 
\label{lingfr}
\ee
where $\O^{ij}(z)$ is the leading component in the harmonic
expansion of the parameter $\O^{--}(z,u)$ in  (\ref{pre-gauge}).
We point out that the linearized super-Weyl tensor $\bW_{\a \b}$ \cite{BdeRdeW}
\bea
{\bar D}_\gd^k \bW_{\a \b} =0~, \qquad D^{\a\b}\bW_{\a\b} ={\bar D}^{\ad \bd} {\bar \bW}_{\ad \bd}
\eea
can be expressed in terms of the gravitational superfield in the form \cite{HST}
\bea
\bW_{\a\b} := {\bar D}^4 D_{\a \b} \bH~, \qquad D_{\a \b} :=D^i_\a D_{\b i}
\eea
and proves to be invariant under the gauge transformations (\ref{lingfr}).

\subsection{Type-I supercurrent}
Given a matter system coupled to the minimal supergravity
multiplet, we define its supercurrent and multiplet of anomalies following \cite{KT}
\be 
{\mathbb J} = \frac{\d S}{\d H} ~, \qquad \quad 
\cT^{++} = \frac{\d S}{\d V_5^{++}}
\ee
with $S$ being the matter action. Here the variational derivatives 
with respect to the supergravity prepotentials 
are defined, in the flat superspace limit,  as follows:
\be
\d S = \int  \rd^4x\, \rd^8\theta \, {\rm d}u \, \d H \,\frac{\d S}{\d H}
+ \int {\rm d} \z^{(-4)}\, \d V_5^{++} \,
\frac{\d S}{\d V_5^{++}}~,
\ee
where the analytic  integration measure is defined by
$ {\rm d} \z^{(-4)} := {\rm d}u \, (D^{-})^4$, with ${\rm d}u$ the usual Haar measure for
$SU(2)$ (see e.g. \cite{GIOS} for more details).
The supercurrent $\mathbb J$ is a real harmonic superfield,
$\breve{\mathbb J} = \mathbb J$, while the multiplet  of anomalies
 $\cT^{++}$ is a real analytic superfield,
\be
D^+_{{\a}} \cT^{++} = {\bar D}^+_\ad \cT^{++}=0~, \qquad\breve{\cT}^{++} = \cT^{++}~.
\ee
The action is required to be invariant under the  gauge transformations
(\ref{pre-gauge}).  This implies that 
\bea
\d_\O S &=&  \frac{1}{4} \int  \rd^4x\, \rd^8\theta \, {\rm d}u \; {\mathbb J}\, (D^+)^2 \O^{--}
+{\rm i} \int  {\rm d} \z^{(-4)} 
\cT^{++} \,(D^+)^4 \O^{--} ~+~{\rm c.c.}\non \\
&=& \int  \rd^4x\, \rd^8\theta \, {\rm d}u \; \O^{--}
\left\{ \frac{1}{4} (D^+)^2 {\mathbb J} + {\rm i}\, \cT^{++} \right\}
~+~{\rm c.c.} = 0 \non
\eea
for arbitrary $\O^{--}$. As a consequence, we get
the conservation equation
\bea
\frac{1}{4} (D^+)^2 {\mathbb J} + {\rm i}\, \cT^{++} &=&0 ~,
\qquad 
\frac{1}{4} ({\bar D}^+)^2 {\mathbb J}  -{\rm i}\, \cT^{++} =0 ~.
\label{conslaw2} 
\eea
Next, the invariance of $S$ with respect to the $l^{--}$ transformations, (\ref{gf2}),  
means
\be
\d_l S =  - \int  \rd^4x\, \rd^8\theta  \, {\rm d}u \; 
(D^{++} l^{--}) \, {\mathbb J }= 
\int  \rd^4x\, \rd^8\theta \, {\rm d}u \; l^{--} D^{++} {\mathbb J} = 0
\non
\ee
for arbitrary $l^{--}$, and hence 
\be 
D^{++} {\mathbb J} = 0~.
\label{aa}
\ee
We see that the matter supercurrent 
is $u$-independent, ${\mathbb J} = \cJ(z)$.
Finally, the action is invariant under the $U(1)$ gauge transformations (\ref{gf3}),
\be
\d_\l S = - \int {\rm d} \z^{(-4)}\,  \cT^{++} \,D^{++}\l = \int {\rm d} \z^{(-4)}\, \l  D^{++}\cT^{++} =0~,
\ee
and hence
\be
D^{++} \cT^{++} = 0\;.
\label{bb}
\ee
The general solution of this equation in the central frame reads
\be
\cT^{++} (z,u) = \cT^{ij}(z) u^+_i u^+_j ~.
\label{trace2}
\ee
Since $\cT^{++} (z,u)$ has to be analytic, 
the multiplet of anomalies  $\cT^{ij}$
satisfies eq. (\ref{trace}). The equations (\ref{conslaw2}), (\ref{aa}) and (\ref{trace2}) imply 
that the conservation law (\ref{sccl}) holds.

If the theory possesses a restricted chiral superfield $X$ constrained by
\bea
{\bar D}_\ad^i X =0 ~, \qquad D^{ij} X = \bar D^{ij} \bar X~,
\eea
then the supercurrent and the anomaly supermultiplet can be modified by adding 
improvement terms \cite{AnB}
\bea
\cJ \to \cJ +i ( \bar X -X )~, \qquad 
\cT^{ij} \to \cT^{ij} + \frac{1}{4} D^{ij} X
\label{improvement}
\eea
without changing the conservation equation (\ref{sccl}).
This is similar to the situation in $\cN=1$ supersymmetric theories \cite{FZ,KS2,K-var}.

Superconformal field theories can be coupled to the Weyl multiplet only, and hence
for such theories
\be
\cT^{ij} =0~. 
\ee
An example of superconformal theories is the 
vector multiplet model (\ref{holomor}) in the case that $F(W^I)$ is a homogeneous function of degree two,
\be
W^I F_I (W) =2 F(W)~.
\ee
Under this condition, the multiplet of anomalies is zero, in accordance with (\ref{holomor2}).

Another example is the improved $\cN=2$ tensor multiplet model\footnote{The harmonic superspace 
formulation  of the improved $\cN=2$ tensor multiplet is given in \cite{GIO}.}
 \cite{deWPV,LR}, which can be written in
projective superspace \cite{KLR,LR-projective1,LR-projective2} as
\begin{align}
S_{\rm IT}=-\frac{1}{2\pi} \oint v^i dv_i \int \rd^4x\, \D^{(-4)} \cG^{(2)} \ln  \cG^{(2)}~, 
\qquad \cG^{(2)}:= \cG^{ij }v_i v_j ~,
\label{ITM}
\end{align}
with the superfield $G^{ij} =G^{ji} $ describing the tensor multiplet,
\bea \label{eq_imptensor}
(\cG^{ij})^* = \cG_{ij} ~, 
\qquad D^{(i}_\a \cG^{jk)} =  {\bar D}^{(i}_\ad \cG^{jk)} = 0~.
\eea
The action involves the following fourth-order differential operator:
\bea
\D^{(-4)} := \frac{1}{16} \nabla^\a \nabla_\a {\bar \nabla}_{ \bd}  {\bar \nabla}^{ \bd} ~, \quad
\nabla_\a := \frac{1 }{ (v,u)} 
{ u_i} D^i_\a 
~, \quad 
{\bar \nabla}_{ \bd} := \frac{1 }{(v,u)} 
u_i{\bar D}^i_{ \bd} ~,~~
\label{2.26}
\eea
where $(v,u):= v^i u_i$.
Here $u_i$ is a fixed isotwistor chosen to be arbitrary modulo 
the condition $(v,u) \neq 0$ along the integration contour.
The supercurrent for the model  (\ref{ITM}) can be shown to be
\begin{align}
\cJ = -\sqrt{\frac{1}{2} \cG^{ij} \cG_{ij}}\equiv -\cG~, \qquad
\cT^{ij} =0~.
\end{align}
The condition that $\cT^{ij}$ vanishes (i.e. that $D^{ij} \cJ = 0$)
follows (after some algebra) from the equation of motion for the tensor multiplet,
which may be written \cite{Siegel:1982wd}
\begin{align}
0 = \frac{\cG}{2} \bar D_{ij} \left(\frac{\cG^{ij}}{\cG^2} \right) = 
\frac{1}{6} \frac{\bar D_{ij} \cG^{ij}}{\cG}
- \frac{1}{9} \bar D_{\dalpha k} \cG^{ki} \bar D^\dalpha_\ell \cG^{\ell j} \frac{\cG_{ij}}{\cG^3}~.
\end{align}

As a final example of a conformal current, we consider an $\cN=2$ 
superconformal model of  interacting tensor and  vector multiplets 
\cite{Siegel:1982wd} given by
\bea
S&=& \frac{1}{2} \int \rd^4x\, \rd^4\theta\, \cW^2
-\frac{1}{2\pi} \oint v^i dv_i \int \rd^4x\, \D^{(-4)} \cG^{(2)} \ln  \cG^{(2)} \non \\
&&\qquad +  \left(\l \int \rd^4x\,\rd^4\theta\, \psi \, \cW + \rm{c.c.}\right)~,
\label{MVTM}
\eea
where $\psi$ is the chiral prepotential of $\cG^{ij}$
\begin{align}
\cG^{ij} = \frac{1}{4} D^{ij} \psi + \rm{c.c.}
\end{align}
and $\lambda$ is a real constant.
The {\it gauge-invariant} action (\ref{MVTM}) describes a superconformal massive  tensor multiplet 
(or equivalently, a massive  vector multiplet).
The interaction term may equally well be
written in projective superspace
\begin{align}
\frac{\lambda}{2\pi} \oint v^i dv_i \int \rd^4x\, \D^{(-4)} \cG^{(2)} \,\cV~,
\end{align}
where $\cV(v^i)$ is the tropical prepotential for  the vector multiplet.\footnote{The  tropical prepotential 
$\cV(v^i)$ is a homogeneous function of $v^i$ of degree zero.}
The field strength $\cW$ 
is given in terms of $\cV$ as 
\begin{align}
\cW = \frac{1}{8\pi} \oint v^i dv_i  \,{\bar \nabla}^2 \cV~,
\end{align}
with ${\bar \nabla}_{ \bd} $ defined in (\ref{2.26}).
In either form the term is topological (i.e. it is independent of the
supergravity prepotential) and so the supercurrent is simply the sum of the
two free supercurrents
\begin{align}
\cJ = \cW \bar \cW -\sqrt{\frac{1}{2} \cG^{ij} \cG_{ij}}~, \qquad
\cT^{ij} =0~.
\end{align}
Demonstrating that $D^{ij} \cJ$ vanishes requires the equations of motion\footnotemark
\begin{subequations}
\begin{align}\label{eq_temp1}
\frac{\cG}{2} \bar D_{ij} \left(\frac{\cG^{ij}}{\cG^2} \right) &= -4\lambda \cW ~,\\
D^{ij} \cW &= -4 \lambda \cG^{ij}~.
\end{align}
\end{subequations}
\footnotetext{The combination appearing on the left side of \eqref{eq_temp1}
must be reduced chiral for the equation to be sensible. This feature was
discussed in \cite{deWPV} and elaborated upon in \cite{Siegel:1982wd}.}

\subsection{Type-II supercurrent}
Many non-superconformal theories must couple to a second supergravity compensator. 
Suppose the latter is  an $\o$-hypermultiplet \cite{GIKOS}, that is a real unconstrained analytic superfield
$\o(z,u)$. It proves to be inert under the $\O$-transformations (\ref{pre-gauge}). 
In the linearized approximation, it changes under the $l$-transformations as
follows\footnote{This rule assumes that the background value of $\omega$ is set
to $-2$. A more general case will be discussed in Appendix \ref{omega_linearized}.}
\be
\d_l \o = - (D^+)^4 D^{--}l^{--}~.
\ee
This transformation law implies that in the present case, the equation (\ref{aa})
must be modified as follows \cite{KT}:
\bea
D^{++} {\mathbb J} + D^{--} \cX^{(+4)} &=& 0~, 
\label{conslaw3}
\eea
where we have introduced 
\bea
\cX^{(+4)}:=\frac{\d S}{\d \o}= \breve{\cX}{}^{(+4)}~, \qquad 
D^+_\a {\cX}^{(+4)} = {\bar D}^+_\ad {\cX}^{(+4)}=0~.
\eea
We see that the supercurrent $\mathbb J$ becomes $u$-dependent.
As to the equation (\ref{conslaw2}), it remains intact.  

In practice,  the superfield $\cX^{(+4)}$ is often characterized by the additional property:
\bea
D^{++} {\cX}^{(4)} =0 \quad \Longrightarrow \quad (D^{++})^3 {\mathbb J} =0~.
\eea
These conditions and the conservation equation  (\ref{conslaw3}) imply that 
\bea
{\mathbb J}(u) = \cJ -2 {\cX}^{ijkl} u^+_i u^+_j u^-_k u^-_l ~, \qquad 
{\cX}^{(+4)} (u) =  {\cX}^{ijkl} u^+_i u^+_j u^+_k u^+_l ~,
\eea
where $ {\cX}^{ijkl} $ obeys the analyticity constraints
\bea
D^{(i}_\a  {\cX}^{jklm)} = {\bar D}^{(i}_\ad  {\cX}^{jklm)}=0~.
\eea   
The conservation equation  (\ref{conslaw2}) turns into 
\bea
\frac{1}{4} {\bar D}^{ij} \cJ =  \frac{1}{20} {\bar D}_{kl} {\cX}^{klij} + {\rm i}\, \cT^{ij}~.
\label{Stelle}
\eea
Setting $\cT^{ij} =0$ gives the supercurrent multiplet introduced by Stelle \cite{Stelle}.

As an example of theories with supercurrent (\ref{Stelle}), we consider the 
massive vector multiplet \cite{GIKOS}
\bea
S_{\rm V}^{(m)} &=& \hf \int \rd^4x\, \rd^4\theta\, W^2 - \hf m^2 
\int {\rm d}\zeta^{(-4)}\,
(V^{++})^2 ~.
\eea
The chiral field strength, $W$,  can be expressed in terms of $V^{++}$  \cite{GIOS,Z} as 
\bea
W(z)= {1\over 4} \int {\rm d}u \, 
({\bar D}^-)^2 \,V^{++}(z,u)~.
\eea
The equation of motion is
\be
{1\over 4} (D^+)^2 W - m^2\, V^{++} =0\quad \Longrightarrow \quad D^{++}V^{++}=0~.
\ee
Here it is worth remembering that the field strength obeys 
the Bianchi identity
 $ (D^+)^2\, W= ({\bar D}^+)^2 {\bar W}$.
The supercurrent multiplet is
\begin{subequations}
\bea
{\mathbb J} &=& W {\bar W} -\hf m^2\, V^{++}(D^{--})^2V^{++}~,\\
{\cX}^{(4)} &=& \hf m^2\, (V^{++})^2~, \qquad {\cT}^{++}=0~. 
\eea
\end{subequations}
The massless case ($m=0$) is both gauge invariant and superconformal.

Another example is the massive tensor multiplet \cite{HST}
\bea
S_{\rm T}^{(m)} &=&  \hf 
\int {\rm d}\zeta^{(-4)}\,(G^{++})^2
-{1\over 4} m^2 \Big\{   \int 
\rd^4x\, \rd^4\theta\,
{ \J}^2 
+{\rm c.c.} \Big\} ~. 
\label{n=2tensor}
\eea
The field strength 
\bea
G^{++}(z,u):= G^{ij} (z) u^{+}_i u^+_j~, \qquad
G^{ij}= \frac{1}{8} D^{ij} { \J} + \frac{1}{8}{\bar D}^{ij} {\bar { \J}} ~, \qquad {\bar D}_\ad^i { { \J}}=0~.
\eea
This action generates the following equation of motion 
\bea
{1 \over 4} ({\bar D}^-)^2 G^{++} 
-m^2 \, { \J}=0~
\eea
and its conjugate.
The supercurrent multiplet is
\begin{subequations}
\bea
{\mathbb J} &=& m^2 { \J} {\bar {\J}} -\hf  G^{++}(D^{--})^2 G^{++}~,\\
{\cX}^{(4)} &=& \hf  (G^{++})^2~, \qquad {\cT}^{++}=0~.
\eea
\end{subequations}
The massless case ($m=0$) is gauge invariant but \emph{not} superconformal.

\subsection{Type-III supercurrent}
We now turn to studying the supercurrent corresponding to  
the off-shell formulation for $\cN=2$ Poincar\'e 
supergravity proposed in \cite{deWPV}. 
At the linearized level, the tensor compensator $\bG^{ij}$ can be represented as above, 
\bea
\bG^{ij}= \frac{1}{4} D^{ij} {\bm \J} +
\frac{1}{4}{\bar D}^{ij} {\bar {\bm \J}} ~, \qquad {\bar D}_\ad^i {\bm \J}=0~.
\label{234}
\eea
It can be shown that the chiral prepotential $\bm \J$ is inert under the $l$-transformations (\ref{gf2}).
On the other hand, the  $\O$-gauge symmetry  (\ref{pre-gauge}) acts on $\bm \J$ as
\be
\d_\O {\bm \J} =  \frac{1}{3} g_{ij} \bar D^4 \bar \Omega^{ij}  ~,
\ee
where $g_{ij}$ denotes the expectation value of the tensor compensator.
This transformation law allows us to read off the supercurrent conservation equation:
\bea
\frac{1}{4} {\bar D}^{ij} \cJ +g^{ij} \cY= {\rm i}\, \cT^{ij} ~,
\label{typeIIIce}
\eea
where we have denoted
\bea
\cY:= \frac{\d S}{\d \bm \J} ~, \qquad {\bar D}_\ad^i {\bm \J}=0~.
\eea
Since the chiral prepotential $\bm \J$ in (\ref{234}) is defined modulo gauge transformations
generated by a restricted chiral superfield, 
the multiplet $\cY$ is a restricted chiral superfield, 
\be
D^{ij} \cY = {\bar D}^{ij} \bar \cY~.
\ee

If the theory possesses  a composite tensor multiplet $L^{ij}$ such that 
\bea
(L^{ij})^* = L_{ij} ~, 
\qquad D^{(i}_\a L^{jk)} =  {\bar D}^{(i}_\ad L^{jk)} = 0~,
\eea
then $\cJ$ and $\cY$ can be modified by adding improvement terms
\bea
\cJ \to \cJ +g_{ij} L^{ij}~, \qquad \cY \to \cY -\frac{1}{12} {\bar D}_{ij} L^{ij}
\eea
without changing the conservation equation (\ref{typeIIIce}). This can be compared with 
eq. (\ref{improvement}) which describes the structure of improvement terms for the type-I supercurrent.

\section{The linearized $\N=2$ supergravity action}
\setcounter{footnote}{0}

Our goal is to linearize the action (\ref{1.11a})--(\ref{1.11c}) around 
Minkowski superspace
which is an exact  solution of the $\cN=2$ Poincar\'e supergravity 
equations. We represent the compensators in the form 
\begin{align}\label{eq_linearize}
\cW = w + \qW, \qquad
\cG^{ij} = g^{ij} + \qG^{ij}
\end{align}
where $w$ and $g^{ij}$ are constant background values
satisfying the equation of motion for the gravitational superfield
\bea
w \bar w = g ~, \qquad g:= \sqrt{\hf g^{ij}g_{ij} }~.
\label{bem}
\eea 
There is no background value for the gravitational superfield.
For the vector compensator, its linearized field strength $\bW$ obeys 
the flat-superspace version of the equations (\ref{1.12}). As to the tensor compensator, 
its linearized field strength $\bG^{ij}$ obeys 
the flat-superspace version of the equations (\ref{1.14}).

The terms involving only the linearized compensators are easy to find:
\bea
&&
-\frac{1}{2} \int \rd^4x\, \rd^4\theta \,\qW \qW + 
\frac{1}{2\pi} \oint v^i dv_i \int \rd^4x\, 
\D^{(-4)}\,
\frac{  \bG^{(2)}  \bG^{(2)} }{2 g^{(2)}} \non \\
&=& -\frac{1}{2} \int \rd^4x\, \rd^4\theta \,\qW \qW
-\frac{i}{8\pi g}  \oint v^i dv_i \int \rd^4x\, 
\D^{(-4)}\,
\frac{  \bG^{(2)}  \bG^{(2)} }{v^{\1} \, v^{\2} } ~.
\label{3.3}
\eea
Here the second  term is given in the projective superspace setting.
We note that 
\bea 
S_{\rm T} = 
\frac{i}{4\pi }  \oint v^i dv_i \int \rd^4x\, 
\D^{(-4)}\,
\frac{  \bG^{(2)}  \bG^{(2)} }{v^{\1} \, v^{\2} } 
\eea
describes a free massless $\cN=2$ tensor multiplet \cite{KLR}.
In appendix A we give a different derivation of the second term in (\ref{3.3}), which is based on the use of 
$\cN=1$ superfields.

It is possible to write the second term in (\ref{3.3}) in harmonic superspace \cite{GIO} as
\begin{align}
-\frac{1}{4g} \int \rd\zeta^{(-4)}\, \qG^{++} \qG^{++}~,
\end{align}
or without any recourse to auxiliary space ${\mathbb C}P^1$  simply
 in the form \cite{SSW}
 \begin{align}
-\frac{1}{320g} \int \rd^4x\, D^{(ij} \bar D^{kl)} (\qG_{ij} \qG_{kl})~,
\end{align}
where the indices $ijkl$ are totally symmetrized (with a factor of $1/4!$ included).
We will use the harmonic form in what follows.

The linearized gravitational  superfield $\qH$ 
and the linearized compensators $\qW$ and $\qG^{ij}$
transform under the supergravity gauge transformations\footnote{As compared with (\ref{lingfr}),  we have rescaled the gauge parameter and switched to bold-face notation,
$\O_{ij} \to 12{\bf \O}_{ij}$.} 
\begin{subequations}
\begin{align}
\delta \qH &=  D^{ij} \qOmega_{ij} + \bar D_{ij} \bar\qOmega^{ij}~, \\
\delta \qW &= - \bar D^4 D^{ij} (\bar w \qOmega_{ij} + w \bar \qOmega_{ij}) ~,\\
\delta \qG_{ij} &= 
D_{ij} \bar D^4 (\bar \qOmega^{kl} g_{kl}) +
     \bar D_{ij} D^4 (\qOmega^{kl} g_{kl})~.
\end{align}
\end{subequations}
The transformation rule for $\qW$ can be derived from \eqref{pre-gauge}.
The rule for $\qG_{ij}$ can similarly be derived from the transformation
of analytic densities considered in \cite{KT}.\footnote{This transformation of
$\qG_{ij}$ was also postulated in \cite{MSW}.}

Using these transformation rules, it is possible to ``complete'' the pure compensator
actions by adding terms involving $\qH$ to make the entire result gauge invariant. However,
it is more illuminating to first motivate the terms linear in $\qH$.
These must arise from the coupling of $\qH$ to the supercurrent:
\begin{align}\label{eq_Noether}
\int \rd^4x\, \rd^8\theta\, \qH \cJ = 
\int \rd^4x\, \rd^8\theta\, \qH (\mathcal G - \mathcal W \bar {\mathcal W})~,
\end{align}
where $\mathcal G = \sqrt{\frac{1}{2} \cG^{ij} \cG_{ij}}$ and $\mathcal W$ involve
the full nonlinear superfields. Expanding to first order using
\eqref{eq_linearize} yields both the background equation of motion, eq. (\ref{bem}),
as well as the second-order terms involving a single $\qH$:
\begin{align}
\int \rd^4x\, \rd^8\theta\, \qH \Big(\frac{1}{2g} g_{ij} \qG^{ij} - w \bar \qW - \bar w \qW \Big)~.
\end{align}

We may then write down the linearized action as
\begin{align}
S_{\rm SUGRA} = S_W + S_G + S_H~,
\end{align}
where
\begin{subequations}
\begin{align}
S_{W} &= -\frac{1}{2} \int \rd^4x\, \rd^4\theta \,\qW \qW 
     - \int \rd^4x\, {\rm d}^8\theta\, (\bar w \qW \qH + w \bar \qW \qH) ~,\\
S_G &= -\frac{1}{4g} \int \rd\zeta^{(-4)} (\qG^{++})^2
     + \frac{1}{2g} \int \rd^4x\, \rd^8\theta\, g_{ij} \,\qG^{ij} \qH
\end{align}
\end{subequations}
are all the terms involving $\qW$ and $\qG^{ij}$, respectively.
The remaining $S_H$ must involve all terms second order in $\qH$.

The explicit form of $S_H$ can be fixed by considering the gauge transformations of 
the functionals $S_W$ and $S_G$:
\begin{subequations}
\begin{align}
\delta S_W &= \int \rd^4x\, \rd^8\theta\, \Big(
     \bar w^2 \qH \bar D^4 D^{ij} \qOmega_{ij}
     + w \bar w \qH  \bar D^4 D^{ij} \bar\qOmega_{ij} + {\rm c.c.}
     \Big)~,
\\
\delta S_G &= \frac{1}{2 g} \int \rd^4x\, \rd^8\theta\, \Big(
     g_{ij} g_{jk} \qH D^{ij} \bar D^4 \bar\qOmega^{kl} + {\rm c.c.}
     \Big)~.
\end{align}
\end{subequations}
These can be cancelled by $\delta S_H$ where
\begin{align}\label{eq_SH}
S_H = \hf \int \rd^4x\, \rd^8\theta\,  \Bigg\{&
     w\bar w \qH \left(\Box 
     -\frac{1}{16} D^{ij} \bar D_{ij} 
        \right) \qH
      -    \bar w^2\, \qH \bar D^4 \qH
     - w^2\, \qH D^4 \qH~~~~~~~~
       \eol
     & \;\;\;\;\;\;\;\;\;
     - \frac{1}{32 g} g_{ij} g_{kl} \, \qH D^{ij} \bar D^{kl} \qH \Bigg\}~.
\end{align}
The second and third terms of $S_H$ are chosen to cancel $\delta S_W$ and
$\delta S_G$; a remaining term is left that can be cancelled by
the first term provided the background equation of motion $w \bar w = g$
holds.\footnote{We have written the coefficient of the first term of $S_H$
as $w \bar w$, but it could just as well be written $g$ or even
$w\bar w/3 + 2 g / 3$ since we are assuming the background to be on-shell.}

It is natural to choose the background  super-Weyl gauge so that 
\be
w \bar w = g = 1~.
\ee
In this gauge $w$ is a pure phase which breaks the background $U(1)_R$ invariance
while $g^{ij}$ is a unit isospin vector which breaks $SU(2)_R$ to a $U(1)$ subgroup.
The linearized action then takes the form
\begin{subequations}
\bea
S_{\rm SUGRA} &=& S_W + S_G + S_H ~, \label{3.15a}\\
S_{W} &=& -\frac{1}{2} \int \rd^4x\, \rd^4\theta \,\qW \qW 
     - \int \rd^4x\, \rd^8\theta\, (\bar w \qW \qH + w \bar \qW \qH) ~,\\
S_G &=& -\frac{1}{4} \int \rd\zeta^{(-4)} (\qG^{++})^2
     + \frac{1}{2} \int \rd^4x\, \rd^8\theta\, g_{ij} \,\qG^{ij} \qH ~, \label{3.15c}\\
S_H &=&\hf  \int \rd^4x\, \rd^8\theta\,  \bigg\{
     \qH \Big( \Box  - \frac{1}{16} D^{ij} \bar D_{ij} 
     \Big) \qH \non \\
     && \qquad -
     \bar w^2\, \qH \bar D^4 \qH
     - w^2\, \qH D^4 \qH
     - \frac{1}{32} g_{ij} g_{kl} \, \qH D^{ij} \bar D^{kl} \qH \bigg\}~.~~~
\label{3.15d}
\eea
\end{subequations}
This linearized supergravity action is one of our main results.

\section{The linearized $\N=2$ supergravity  action in terms of $\N=1$ superfields}
To better understand the physics of the linearized action, we consider
its reduction to $\N=1$ superfields by performing all Grassmann integrals
involving $\theta_{\ul 2}$ and $\bar\theta^{\ul 2}$. This will leave manifest
one supersymmetry (involving $\theta \equiv \theta_{\ul 1}$ and
$\btheta \equiv \btheta^{\ul 1}$). The precise choice of which supersymmetry
to leave manifest is not physical: different choices involve an $SU(2)_R$ rotation
of the background and so involve different choices for the isospin unit vector
$g_{ij}$.

\subsection{Setup}
For the pure Maxwell compensator action, this calculation is straightforward. 
We start from
\begin{align}
-\frac{1}{2} \int \rd^4x\, \rd^4\theta\, \qW \qW~,
\end{align}
where $\qW$ obeys the Bianchi identity
\begin{align}
\label{eq_BianchiW2}
D_{ij} \qW = \bar D_{ij} \bar \qW~.
\end{align}
This can be easily rewritten in terms of $\N=1$ superfields (making
use of the Bianchi identities) as
\begin{align}
\label{4.3}
-\int \rd^4x\, \rd^2\theta\, W^\alpha W_\alpha
- \int \rd^4x\, \rd^4\theta\, \c \bar \c~,
\end{align}
where
\begin{subequations}
\begin{gather}
\c := \qW\vert, \;\;\;
\bar\c := \bar\qW\vert ~, \\ 
\label{eq_Walpha}
W_\alpha := -\frac{i}{2} D_\alpha{}^{\ul 2} \qW\vert, \;\;\;
{\bar W}_\dalpha := +\frac{i}{2} \bar D_\dalpha{}_{\ul 2} \bar \qW\vert, \;\;\;
\end{gather}
\end{subequations}
are $\N=1$ chiral and antichiral superfields, respectively,
and $\vert$ denotes taking $\theta_{\ul 2} = \bar\theta^{\ul 2} = 0$.
The factor of $i$ in \eqref{eq_Walpha} 
is chosen so that \eqref{eq_BianchiW2}
implies the $\N=1$ Bianchi identity
\begin{align}
D^\alpha W_\alpha = \bar D_\dalpha \bar W^\dalpha~.
\end{align}
Note that the $\N=1$ actions in (\ref{4.3}) both have the wrong sign, implying that
both $\N=1$ fields will play the role of compensators.

${}$For the pure tensor compensator action written in harmonic superspace, we have
\begin{align}
-\frac{1}{4} \int \rd\zeta^{(-4)}\, (\qG^{++})^2 ~, \qquad  \qG^{++} = \qG^{ij} u_i^+ u_j^+~,
\end{align}
where $\qG^{++} $ obeys the constraint
$D_\alpha^+ \qG^{++} = D_\dalpha^+ \qG^{++} = 0$. These constraints imply that
the components $\qG^{ij}$ consist of an $\N=1$ tensor multiplet $G$
(also known as a real linear multiplet) and an $\N=1$ chiral scalar multiplet $\eta$:
\begin{gather}
\eta := \qG_{\1\1}{}\vert ~, \qquad
\bar \eta := \qG_{\2\2}{}\vert~,\qquad
G = +2i \qG_{\1\2}{}\vert~.
\end{gather}
This is easily rewritten in terms of $\N=1$ superfields as
\begin{align}
-\hf \int \rd^4x\, \rd^4\theta\, \Big(
\eta \bar \eta 
-\hf G^2 \Big)~.
\end{align}
We note again the wrong signs, which is an indicator that these are
compensators.

The remainder of the terms to be reduced to $\N=1$ involve the $\N=2$
superfield $\qH$. Because of the gauge freedom for $\qH$, most of
its $\N=1$ content is pure gauge. It is advantageous to eliminate
as much of the gauge degrees of freedom as possible to identify the physical
$\N=1$ superfields. The simplest gauge choice is a Wess-Zumino gauge,
where we use the $\theta^{\ul 2}$ and $\btheta_{\ul 2}$ components of
$\qOmega_{ij}$ to fix the lower components of $\qH$ to zero:
\begin{align}\label{eq_WZgaugeH}
\qH\vert = D_\alpha^{\ul 2} \qH\vert = 
     D_\dalpha{}_{\ul 2} \qH\vert = (D^{\ul 2})^2 \qH\vert = 
     (\bar D_{\ul 2})^2 \qH\vert = 0~.
\end{align}
The remaining higher components of $\qH$ are identified
as\footnote{The precise definitions of the higher components of
$\qH$ are ambiguous up to terms that vanish in Wess-Zumino gauge.}
\begin{subequations}\label{eq_Hcomps}
\begin{align}
H_{\alpha \dalpha} &:= \frac{1}{4} [D_\alpha{}^{\ul 2}, \bar D_{\dalpha}{}_{\ul 2}] \qH\vert ~,\\
\Psi_\alpha & := \frac{1}{8} (\bar D_{\ul 2})^2 D_\alpha{}^{\ul 2} \qH\vert  ~,\\
U & := \frac{1}{16} D^\alpha{}^{\ul 2} (\bar D_{\ul 2})^2 D_\alpha{}^{\ul 2} \qH\vert~.
\end{align}
\end{subequations}
Here $H_{\alpha \dalpha}$ is the $\N=1$ supergravity multiplet,
$\Psi_\alpha$ is the gravitino matter multiplet associated with the second
gravitino, and $U$ is a real auxiliary superfield.

As usual when imposing Wess-Zumino gauge, a residual gauge transformation
remains. In the case under consideration, that invariance is
\begin{subequations}\label{eq_N1gauge}
\begin{align}
\delta H_{\alpha \dalpha} &= D_\alpha{\bar L}_\dalpha - \bar D_\dalpha L_\alpha~, \\
\delta \Psi_\alpha &= D_\alpha \Omega + \Lambda_\alpha ~, 
\qquad \qquad \qquad \qquad \qquad {\bar D}_\ad \L_\a =0~,\\
\delta U &= \Phi + \bar \Phi
     - \frac{i}{2} \partial^{\dalpha \alpha} ({\bar D}_\dalpha L_\alpha + D_\alpha {\bar L}_\dalpha)~,
     \qquad {\bar D}_\ad \F =0~,
\end{align}
\end{subequations}
where $\Phi$ and $\Lambda_\alpha$ are chiral and $L_\alpha$ and $\Omega$ are
unconstrained complex superfields. These gauge parameters may be defined
in terms of complicated spinorial derivatives of $\qOmega_{ij}$ and $\bar\qOmega^{ij}$.
The details are given in Appendix \ref{N1gauge}.

Because the $\N=2$ compensators transform under $\qOmega_{ij}$,
their $\N=1$ descendants should transform under the residual transformations
\eqref{eq_N1gauge}. For the Maxwell multiplet compensators, one finds
\begin{subequations}
\begin{align}
\delta \chi &= -w \Phi - \frac{1}{4} w \bar D^2 D^\alpha L_\alpha ~,\\
\delta W_\alpha &= \frac{1}{4} \bar D^2 D_\alpha \left(i \bar w \Omega - i w \bar\Omega \right)~.
\end{align}
\end{subequations}
Note that $\chi$ transforms as a chiral compensator for the $\N=1$
supergravity sector while $W_\alpha$ transforms as a chiral spinor compensator
for the gravitino multiplet $\Psi_\alpha$.

The tensor sector is more intricate:
\begin{subequations}
\bea
\delta G &=& 2i \delta \qG_{\1\2}\vert
     = \frac{i}{2}  g_{\1\2} (D^\alpha \bar D^2 L_\alpha + \bar D_\dalpha D^2 L^\dalpha)
     - i g_{\1\1} D^\alpha \Lambda_\alpha
     + i g_{\2\2} \bar D_\dalpha \bar\Lambda^\dalpha ~, ~~~~~~\\
\delta \eta &=& \delta \qG_{\1\1}\vert = -g_{\1\2} \bar D^2 \bar\Omega
     + g_{\1\1} \Phi ~,\\
\delta \bar \eta &=& \delta  \qG_{\2\2}\vert = + g_{\1\2} D^2 \Omega
     + g_{\2\2} \Phi~.
\eea
\end{subequations}
The linear superfield $G$ transforms as a linear compensator
for supergravity only if $g_{\1\2}$ is nonzero and as a linear compensator for
the gravitino only if $g_{\1\1}$ (and its conjugate $g_{\2\2}$) are nonzero. Similarly,
$\eta$ is a chiral compensator for the gravitino only if $g_{\1\2}$ is nonzero.

The vacuum value of $g_{ij}$ not only breaks $SU(2)_R$ to some $U(1)$ subgroup
but also strongly affects the form of the $\N=1$ supergravity sector, as the
structure of the supergravity section depends greatly on these
parameters. We report the relevant formulae for arbitrary $g_{ij}$ in Appendix
\ref{generalN1reduction}; here, we will analyze in detail only the two interesting
simple cases. The first, which we call case I, involves taking $g_{\1\1} = g_{\2\2} = 0$;
the second, case II, is $g_{\1\2} = 0$.

\subsection{Case I: New minimal supergravity}

In this case, $g_{\1\1} = g_{\2\2} = 0$ and $g_{\1\2} =\pm i$. 
This choice is invariant under the diagonal $U(1)$ subgroup of $SU(2)_R$ and so
this $U(1)$ should be manifest in the vacuum of the $\N=1$ reduction.
Moreover, $G$ is a linear compensator for supergravity with no
compensating transformation for the gravitino. Similarly, $\eta$ 
is solely a compensator for the gravitino. This strongly implies that the
supergravity reduction of this model should resemble new minimal supergravity \cite{new}.

Performing the reduction with only $g_{\1\2}=g_{\2\1}$ nonzero 
gives $S = S_W + S_G + S_H$ with
\begin{subequations}
\bea
\label{eq_SW}
S_W &= &- \int \rd^4x\, \rd^2\theta\, W^\alpha W_\alpha - \int \rd^4x\, \rd^4\theta\, 
\Bigg\{
\chi \bar \chi   + 2 i\bar w \Psi^\alpha W_\alpha
     - 2 i w \bar\Psi_\dalpha \bar W^\dalpha
\non \\
    & & \quad  
    +  U (w \bar \chi+ \bar w \chi)
     - \frac{i}{2}  (\bar w \chi - w \bar \chi) \pa_{\a\ad}H^{\ad \a}    \Bigg\}~,~~~~
  \\
S_G &=& \frac{1}{4}  \int \rd^4x\, \rd^4\theta\, \Bigg\{
G^2
 + 2i g_{\1\2} U G  
 - \frac{i}{2}        g_{\1\2 } 
 [D_\a, {\bar D}_\ad ] H^{\ad \a}  G
 \non \\
&& \qquad  - 2 \eta {\bar \eta}    
     - 2 {g_{\1\2}} \big( \Psi^\alpha D_\alpha \eta
       -{\bar \Psi}_\dalpha \bar D^\dalpha {\bar\eta} \big) \Bigg\}~, \\
S_H &=& \int \rd^4x\, \rd^4\theta\, \Bigg\{
      -\frac{1}{16} H^{\dalpha \alpha} D^\beta \bar D^2 D_\beta H_{\alpha \dalpha}
     - \frac{1}{4}  (\partial_{\a\ad}  H^{\ad \a})^2
       +\frac{1}{32} ([D_\a, {\bar D}_\ad ] H^{\ad \a})^2
     \non \\
     && \qquad - \Psi^\alpha \bar D_\dalpha D_\alpha \bar\Psi^\dalpha
     - \frac{\bar w^2}{4} \Psi^\alpha \bar D^2 \Psi_\alpha
     - \frac{w^2}{4} \bar\Psi_\dalpha D^2 \bar\Psi^\dalpha \non \\
     && \qquad - \frac{1}{4} U 
      [D_\a, {\bar D}_\ad ] H^{\ad \a}
     - \frac{1}{2} U^2 \Bigg\}~.
\eea
\end{subequations}

Because $U$ appears in the action without derivatives,
it plays the role of an
$\N=1$ auxiliary superfield. If integrated out, the action becomes the sum
of two decoupled sectors $S = S_{\rm SG} + S_\Psi$.
The $\N=1$ supergravity sector is contained in $S_{\rm SG}$:
\bea
S_{\rm SG} &=& \int \rd^4x\, \rd^4\theta\,
     \Bigg\{-\frac{1}{16} H^{\dalpha \alpha} D^\beta \bar D^2 D_\beta H_{\alpha \dalpha}
     - \frac{1}{4} (\partial_{\a \ad}  H^{\ad \a})^2 
     + \frac{1}{16} ([D_\a, {\bar D}_\ad ] H^{\ad \a})^2 \non \\
     &&\qquad 
      -\hf L\, [D_\a, {\bar D}_\ad ] H^{\ad \a}
      + \frac{3}{2} L^2 \Bigg\}~,
\eea
where we have rescaled the linear compensator $G$ to $L = i g_{\1\2} G / 2$ to make contact
with the conventional normalization of new minimal supergravity (see, e.g., 
\cite{BK} for a review). This action has the gauge invariance
\begin{subequations}
\begin{align}
\delta H_{\alpha \dalpha} &= D_\alpha {\bar L}_\dalpha - \bar D_\dalpha L_\alpha ~,\\
\delta L &= \frac{1}{4} D^\alpha \bar D^2 L_\alpha + \frac{1}{4} \bar D_\dalpha D^2 L^\dalpha ~.
\end{align}
\end{subequations}

The gravitino sector is
\bea
S_\Psi &=& \int \rd^4x\, \rd^4\theta\, \Bigg\{
      -  \Psi^\alpha \bar D_\dalpha D_\alpha \bar\Psi^\dalpha
     - \frac{\bar w^2}{4} \Psi^\alpha \bar D^2 \Psi_\alpha
     - \frac{w^2}{4} \bar\Psi_\dalpha D^2 \bar\Psi^\dalpha 
     \non \\
     &&
     - \frac{g_{\1\2}}{2} \big(  \Psi^\alpha D_\alpha \eta
      -{\bar \Psi}_\dalpha \bar D^\dalpha {\bar \eta} \big)
      - \frac{1}{2} \eta \bar{\eta}     
     - 2i \bar w \Psi^\alpha W_\alpha
     + 2i w \bar\Psi_\dalpha \bar W^\dalpha
    \Bigg\}
       \non \\
  &&   - \int \rd^4x\, \rd^2\theta\,  W^\alpha W_\alpha
\eea
and involves two compensators -- the gaugino field strength $W_\alpha$ and
the chiral superfield $\eta$ -- as well as the phase $w$ and the
imaginary constant $g_{\1\2} =\pm i$. 
If we make the choice $w = -i$ and perform the field redefinition
\begin{align}
\phi := - g_{\1\2} \,\eta~, \qquad 
\bar\phi := + g_{\1\2}\, \bar\eta~,
\end{align}
we end up with the massless gravitino action \cite{GS80}
\begin{align}
S_\Psi = \int \rd^4x\, \rd^4\theta\, \Bigg\{
     & - \Psi^\alpha \bar D_\dalpha D_\alpha \bar\Psi^\dalpha
     + \frac{1}{4} \Psi^\alpha \bar D^2 \Psi_\alpha
     + \frac{1}{4} \bar\Psi_\dalpha D^2 \bar\Psi^\dalpha \eol
     &
     + \frac{1}{2} \Psi^\alpha D_\alpha \phi
     + \frac{1}{2} \bar\Psi_\dalpha \bar D^\dalpha \bar\phi
     - \frac{1}{2} \phi \bar\phi
     + 2\Psi^\alpha W_\alpha
     + 2\bar\Psi_\dalpha \bar W^\dalpha
    \Bigg\}\eol
     - \int \rd^4x\, \rd^2\theta\, & W^\alpha W_\alpha~.
     \label{gravitino1}
\end{align}

This action is invariant under
\begin{align}
\delta \Psi_\alpha &= D_\alpha \Omega + \Lambda_\alpha~, 
\qquad  \qquad {\bar D}_\ad \L_\a =0~,\\
\delta W_\alpha &= -\frac{1}{4} \bar D^2 D_\alpha \left(\Omega + \bar\Omega \right) ~,\\
\delta \phi &= -\bar D^2 \bar\Omega~,
\end{align}
with $\Lambda_\alpha$ chiral and $\Omega$ complex unconstrained.
It is possible to remove one or both of the compensators by exhausting
some of the gauge freedom.  
The gravitino model (\ref{gravitino1})
 can be shown to be  equivalent to the Fradkin-Vasiliev-de Wit-van Holten 
 formulation \cite{FV,dWvH} 
for a massless gravitino multiplet, 
derived originally in components.
The above realization,  eq. (\ref{gravitino1}), 
for the gravitino action  is reviewed  in textbooks \cite{GGRS,BK}.

Note that the chiral compensator $\chi$ has completely dropped out of the
action, appearing neither in $S_{SG}$ nor $S_\Psi$. It is a pure gauge degree of
freedom, corresponding to the gauge parameter $\Phi$.

\subsection{Case II: Old minimal supergravity}
For the second case, we have $g_{\1\1} = \gamma$ and $g_{\2\2} = \bar \gamma$
for some $\gamma$ such that $\g \bar \g =1$, 
while $g_{\1\2}=0$. A particular choice of $\gamma$ breaks the
diagonal $U(1)$ subgroup of $SU(2)_R$ (while maintaining some other $U(1)$ subgroup)
and so no manifest $U(1)_R$-symmetry exists
in the vacuum of II. For this choice, the linear compensator $G$
compensates only for the gravitino.

Performing the $\N=1$ reduction gives the same $S_W$ as in \eqref{eq_SW},
while $S_G$ and $S_H$ are altered:
\begin{subequations}
\bea
S_G &= & \frac{1}{4} \int \rd^4x\, \rd^4\theta\, \Bigg\{ G^2 -2 \eta {\bar \eta}
          + {i} (\bar\gamma \eta - \gamma {\bar\eta}) \partial_{\a\ad} H^{\ad \a}     \eol
&& \qquad 
     + 2 U \left(
     \gamma {\bar\eta} + \bar\gamma \eta \right)
     - 2 {i \gamma } \Psi^\alpha D_\alpha G
     + 2 {i \bar\gamma  } \bar\Psi_\dalpha \bar D^\dalpha G \Bigg\} ~,\\
S_H &=&\int \rd^4x\, \rd^4\theta\, \Bigg\{
- \frac{1}{16} H^{\dalpha \alpha} D^\beta \bar D^2 D_\beta H_{\alpha \dalpha}
     - \frac{1}{4} (\partial_{\a\ad} H^{\ad\a})^2      
     + \frac{1}{64} 
       ([D_\a, {\bar D}_\ad ] H^{\ad \a})^2
      \eol
&& \qquad 
     - \Psi^\alpha \bar D_\dalpha D_\alpha \bar\Psi^\dalpha
     - \frac{w^2}{4} \bar\Psi_\dalpha D^2 \bar\Psi^\dalpha
     - \frac{\bar w^2}{4} \Psi^\alpha \bar D^2 \Psi_\alpha
     - \frac{1}{4} \left(\gamma D^\alpha \Psi_\alpha - \bar \gamma \bar D_\dalpha \bar\Psi^\dalpha \right)^2  
     \non \\
   &&    
 \qquad     - \frac{1}{8} U           [D_\a, {\bar D}_\ad ] H^{\ad \a}
  - \frac{3}{4} U^2
     \Bigg\}~.
\eea
\end{subequations}

Integrating out $U$, we find again the sum of two actions: $S = S_{\rm SG} + S_\Psi$.
In this case, the supergravity action $S_{SG}$ is that for old minimal supergravity \cite{old}
\bea
S_{\rm SG} &=& \int \rd^4x\, \rd^4\theta\, \Bigg\{
     - \frac{1}{16} H^{\dalpha \alpha} D^\beta \bar D^2 D_\beta H_{\alpha \dalpha}
     - \frac{1}{4} (\partial_{\a\ad} H^{\ad \a})^2
     + \frac{1}{48} 
        ([D_\a, {\bar D}_\ad ] H^{\ad \a})^2 \non \\\
&&  \qquad    + i (\partial_{\a\ad} H^{\ad \a}) (\sigma - \bar\sigma)
     - 3 \sigma \bar \sigma\Bigg\}~,
\eea
where the chiral compensator $\sigma$ is defined by
\begin{align}
\sigma := \frac{1}{3} \bar w \chi + \frac{1}{3} \bar\gamma \eta~.
\end{align}
The action is gauge invariant under
\begin{subequations}
\begin{align}
\delta H_{\alpha \dalpha} &= D_\alpha {\bar L}_\dalpha - \bar D_\dalpha L_\alpha ~,\\
\delta \sigma &= -\frac{1}{12} \bar D^2 D^\alpha L_\alpha~.
\end{align}
\end{subequations}
Note that $\sigma$ is the particular combination of $\chi$ and $\eta$
which does not transform under the $\Phi$ gauge transformation.
The other linearly independent combination of $\chi$ and $\eta$ -- which
does depend on $\Phi$ -- has dropped out of the action.

The gravitino action is
\begin{align}
S_\Psi = \int \rd^4x\, \rd^4\theta\, \Bigg\{&
     - \Psi^\alpha \bar D_\dalpha D_\alpha \bar\Psi^\dalpha
     - \frac{w^2}{4} \bar\Psi_\dalpha D^2 \bar\Psi^\dalpha
     - \frac{\bar w^2}{4} \Psi^\alpha \bar D^2 \Psi_\alpha \eol
     & - \frac{1}{4} \left(\gamma D^\alpha \Psi_\alpha - \bar \gamma \bar D_\dalpha \bar\Psi^\dalpha \right)^2
     - 2i \bar w \Psi^\alpha W_\alpha
      + 2i w \Psi_\dalpha \bar W^\dalpha \eol
     & - \frac{i}{2} \gamma \Psi^\alpha D_\alpha G
     + \frac{i}{2} \bar\gamma\bar\Psi_\dalpha \bar D^\dalpha G
      + \frac{1}{4} G^2 \Bigg\}\eol
- \int \rd^4x\, \rd^2\theta\,& W^\alpha W_\alpha
\end{align}
and involves again two compensators -- $W_\alpha$ and the linear superfield $G$ --
as well as the constant phase factors $w$ and $\gamma$. If we make the choices
$w = \gamma = -i$, we find
the following gravitino action \cite{LR}:
\begin{align}
S_\Psi = \int \rd^4x\, \rd^4\theta\, \Bigg\{&
     - \Psi^\alpha \bar D_\dalpha D_\alpha \bar\Psi^\dalpha
     + \frac{1}{4} \bar\Psi_\dalpha D^2 \bar\Psi^\dalpha
     + \frac{1}{4} \Psi^\alpha \bar D^2 \Psi_\alpha \eol
     & + \frac{1}{4} \left(D^\alpha \Psi_\alpha + \bar D_\dalpha \bar\Psi^\dalpha \right)^2
     + 2\Psi^\alpha W_\alpha
      + 2\Psi_\dalpha \bar W^\dalpha \eol
     & - \frac{1}{2} \Psi^\alpha D_\alpha G
     - \frac{1}{2} \bar\Psi_\dalpha \bar D^\dalpha G
      + \frac{1}{4} G^2 \Bigg\}\eol
- \int \rd^4x\, \rd^2\theta\,& W^\alpha W_\alpha~.
 \label{gravitino2}
\end{align}
Its  gauge invariance is
\begin{subequations}
\begin{align}
\delta \Psi_\alpha &= D_\alpha \Omega + \Lambda_\alpha ~, \\
\delta W_\alpha &= -\frac{1}{4} \bar D^2 D_\alpha \left(\Omega + \bar\Omega \right)~, \\
\delta G &= - \big( D^\alpha \Lambda_\alpha + \bar D_\dalpha \bar\Lambda^\dalpha \big)~.
\end{align}
\end{subequations}
The gravitino actions  (\ref{gravitino1})  and  (\ref{gravitino2}) are dual to each other \cite{LR}.
This duality is an example of the Legendre transformation between the tensor and the chiral multiplets.

As before, it is possible  to remove the compensators algebraically by exhausting
some of the gauge freedom. In the gauge $W_\a =
 G=0$,  (\ref{gravitino2}) reduces to the  gravitino action 
 discovered by Ogievetsky and Sokatchev \cite{Ogievetsky:1976qb}.
In accordance with the above discussion, it can be considered 
to be dual to the Fradkin-Vasiliev-de Wit-van Holten gravitino action.

\section{Discussion}
\setcounter{footnote}{0}
We have succeeded in constructing the linearized $\N=2$ supergravity
action involving vector and tensor compensators along with the scalar
supergravity prepotential $H$. Such an action was ruled out in \cite{GS82}
where the spinor prepotential $\Psi_{\alpha}^i$ was used for the $\N=2$
supergravity sector. It was argued in \cite{KT} that parametrizations
involving $\Psi_{\alpha}^i$ and those involving $H$ are merely different
choices for an underlying gauge symmetry of a harmonic prepotential;
when a theory is written with the spinor prepotential, it ought then 
to appear only in the combination
\begin{align}\label{eq_HPsi}
H = D^{\alpha}_i \Psi_{\alpha}^i + \bar D_{\dalpha}^i \bar\Psi^{\dalpha}_i
\end{align}
In \cite{GS82} it was argued that while this holds for most terms in
the linearized action, the coupling of the tensor compensator to
$\Psi_\alpha^i$ requires a different form
\begin{align}
\int \rd^4x \,\rd^8\theta\, \qG_{ij} D^{\alpha i} \Psi_\alpha^j + {\rm c.c.}
\end{align}
where $\Psi_\alpha^i$ appears explicitly. This is necessary \emph{only} if
the vacuum is required to respect $SU(2)_R$ invariance.
Several years later, when the 
minimal formulation for $\cN=2$ Poincar\'e supergravity
with a tensor compensator was finally constructed
\cite{deWPV}, it became
clear that the compensator's field strength $\cG_{ij}$ itself, rather than its
chiral prepotential, must gain a vacuum value, and so the theory will
necessarily break $SU(2)_R$ in the appearance of that vacuum value $g_{ij}$.
This allows the parametrization independent coupling
\begin{align}
\int \rd^4x \,\rd^8\theta\, g^{ij} \qG_{ij} \qH
\end{align}
which we have used in this paper and which follows from linearizing the
supercurrent. One may still choose to parametrize $\qH$ via \eqref{eq_HPsi},
but this is not a necessity.

The $\N=1$ reduction of the theory is especially interesting. Once the
$\N=1$ auxiliary superfield $U$ is integrated out (so that the second
supersymmetry is realized only on shell), the two simple choices for $g_{ij}$
reduce to decoupled actions of supergravity and the gravitino multiplet. The
first choice, with $g_{\ul{11}} = g_{\ul{22}} = 0$, corresponds to new minimal
supergravity, where the $\N=1$ conformal supergravity prepotential
couples to a linear compensator. The second choice, with $g_{\ul{12}} = 0$,
corresponds to old minimal supergravity with a chiral compensator.
As is well known, these two theories are dual to each other, but here we
see them appear as different vacua of a single theory.
The same curious features apply to the gravitino sector. In the first case, we
have essentially the Fradkin-Vasiliev-de Wit-van Holten formulation with a
chiral compensator, while in the second case, the Ogievetsky-Sokatchev formulation
with a linear compensator. These gravitino models are well known to be dual to
each other, but here we see them arise as physically equivalent theories, simply
with different vacua.\footnote{This behavior was briefly speculated about in the closing
of \cite{deWPV}.} More intriguing still, the chiral and linear compensators are
traded between the supergravity and gravitino sectors when the background is changed
by an isospin rotation.

Our model for the linearized $\cN=2$ supergravity  (\ref{3.15a})--(\ref{3.15d})
admits several dual formulations obtained by applying superfield Legendre transformations. 
Within the harmonic superspace approach \cite{GIKOS,GIOS}, 
the tensor compensator can be dualized into a $q^+$-hypermultiplet or into an $\o$-hypermultiplet 
using the procedure described in \cite{GIO,GIOS}. For this the tensor multiplet part
(\ref{3.15c}) of the supergravity action should be rewritten in the form
\bea
S_G &=& -\frac{1}{4} \int \rd\zeta^{(-4)} \,\qG^{++}   \Big\{ \qG^{++}
 - 6  g^{--}  (D^+)^4\qH
     - 4 D^{++} \qomega \Big\} ~.
 \eea 
Here $\qomega$ is an unconstrained analytic superfield acting as a Lagrange
multiplier to enforce the constraint $D^{++} \qG^{++} = 0$. One may instead
integrate out $\qG^{++}$ to arrive at a dual action in terms of $\qomega$:
\begin{subequations}
\begin{align}
S_{\rm SUGRA} &= S_W + S_\omega + S_H ~,\\
S_{W} &= -\frac{1}{2} \int \rd^4x\, \rd^4\theta \,\qW \qW 
     - \int \rd^4x\, \rd^8\theta\, (\bar w \qW \qH + w \bar \qW \qH) ~,\\
S_{\omega} &=
     \int \rd\zeta^{-4} D^{++} \qomega D^{++} \qomega
     - 6 \int \rd u\, \rd^4x\, \rd^8\theta \, \omega^{-+} \qH \qomega ~,\\
S_H &= \hf  \int \rd^4x\, \rd^8\theta\,  \bigg\{
     \qH \Big( \Box  - \frac{1}{10} D^{ij} \bar D_{ij} 
     \Big) \qH \non \\
     & \qquad -
     \bar w^2\, \qH \bar D^4 \qH
     - w^2\, \qH D^4 \qH
     + \frac{1}{40} \omega_{ij} \omega_{kl} \, \qH D^{ij} \bar D^{kl} \qH \bigg\}
\end{align}
\end{subequations}
where $\omega^{ij}$ is a unit isospin vector and we work in the gauge where
$w \bar w = \omega^{ij} \omega_{ij} / 2 = 1$.
Because the full nonlinear coupling of an $\omega$-hypermultiplet to
supergravity is known, one may compare the result from the duality transformation
to the result from linearizing the $\omega$-hypermultiplet action explicitly.
The details are given in Appendix \ref{omega_linearized}.

 In the projective superspace approach \cite{KLR,LR-projective1,LR-projective2}, on the other hand, 
 the tensor compensator can be dualized into a polar hypermultiplet (following the terminology 
 of \cite{G-RRWLvU}). The dual formulations thus derived may  lead 
 to new variant $\cN=2$ supercurrent multiplets.

Building on  the linearized $\cN=2$ supergravity  action (\ref{3.15a})--(\ref{3.15d}) and its dual versions, 
an open problem is to  construct their massive extensions.
It would be interesting to understand how the known off-shell realizations for massive 
$\cN=1$ gravitino and supergravity multiplets are imbedded in such $\cN=2$ models
\cite{OS77,AB,BGLP,Zinoviev,GSS,BGKP,GKT-M}.

One can consistently incorporate a cosmological term into the minimal $\cN=2$ supergravity 
with tensor compensator \cite{deWPV}.
In superspace, its description  is achieved  by replacing the action (\ref{1.11c}) with
the following \cite{K-dual08}:
\bea
S^{(m)}_{\rm tensor}&=& 
\frac{1}{2\pi \k^2} \oint_C  v^i \rd v_i
\int \rd^4 x \,{\rm d}^4\q {\rm d}^4{\bar \q}
\,\frac{E}{S^{(2)} \breve{S}^{(2)}}\,
{\cG}^{(2)} \Big\{ 
\ln \frac{{\cG}^{(2)}}{{\rm i}\breve{ \U}^{(1)}  \U^{(1)}} - m \cV\Big\} ~,~~~
\label{5.2}
\eea
Here $m$ is the  cosmological constant, 
and $\cV (v^i)$ is the weight-zero tropical gauge prepotential 
for the vector compensator.
The equation of motion for the vector compensator in this theory is 
\bea
 \Big( \frac{1}{ 4}\cD^{\a(i}\cD_\a^{j)}+S^{ij}\Big) \cW +m \cG^{ij} =0~.
\eea
The equation for the gravitational field (\ref{1.16}) does not change, for the vector-tensor coupling in (\ref{5.2}) 
is topological.
In a super-Weyl and local $U(1)$ gauge $\cW=1$, the above equation reduces to 
$S^{ij} +m \cG^{ij}=0$, and thus ${\bar S}^{ij}=S^{ij}$. 
A maximally symmetric solution in $\cN=2$ supergravity with cosmological term 
 is $\cN=2$ anti-de Sitter superspace
for  which the covariant derivatives $\cD_A = (\cD_a , \cD_\a^i, {\bar \cD}^\ad_i)$
are characterized by the algebra (see \cite{KT-M-ads} for more details):
\begin{subequations}
\bea\{\cD_\a^i,\cD_\b^j\}&=&
4{\bm S}^{ij }M_{\a\b}
+2 \ve_{\a\b}\ve^{ij} {\bm S}^{kl}J_{kl}~,
\qquad
\{\cD_\a^i,\cDB^\bd_j\}=
-2\ri\d^i_j(\s^c)_\a{}^\bd\cD_c
~,~~~
\label{AdS-N2-1}
\\
{[}\cD_a,\cD_\b^j{]}&=&
{\ri\over 2} ({\s}_a)_{\b\gd} {\bm S}^{jk}\cDB^\gd_k~,
\qquad \qquad \qquad \quad ~~
[\cD_a,\cD_b]= - {\bm S}^2
M_{ab}~,
\label{AdS-N2-2}
\eea
\end{subequations} 
with ${\bm S}^2 :=\hf  {\bm S}^{ij}{\bm S}_{ij}$,  and $M_{\a \b} $ and $J_{kl}$ 
the Lorentz and $SU(2)$ generators, respectively. 
Here the  covariantly constant torsion 
${\bm S}^{ij} = \bar {\bm S}^{ij}$ is the background value of $\cS^{ij}$.
It would be interesting to construct a linearized $\cN=2$ supergravity action around the 
anti-de Stitter background.
\\

\noindent
{\bf Acknowledgements:}\\
This work  is supported in part by the Australian Research Council 
and by a UWA Research Development Award.

 \appendix

\section{The improved $\cN=2$ tensor multiplet in $\cN=1$ superspace}
\setcounter{footnote}{0}

Historically, the first formulation for the improved $\cN=2$ tensor multiplet was given using 
$\cN=1$ superfields \cite{LR},  and the component ($\cN=0$) formulation of \cite{deWPV} 
appeared shortly after.\footnote{Ref. \cite{LR} was submitted to the journal Nuclear Physics B
one day earlier than  \cite{deWPV}.} Here we briefly review some aspects of the $\cN=1$ formulation \cite{LR}.

The manifestly $\cN=2$ superconformal action for the 
improved  tensor multiplet model (\ref{ITM})
 can be rewritten in terms of 
 $\cN=1$  superfields. The resulting action  \cite{LR} is
\bea
S_{\rm IT}    =    \int  {\rm d}^4 x \,{\rm d}^4\q \, L_{\rm IT}~, \qquad
 L_{\rm IT}=   \sqrt{ \mathsf{G}^2 +4\vf \bar \vf }
    - \mathsf{G} \, \ln \big( \mathsf{G}+ \sqrt{ \mathsf{G}^2 +4\vf \bar \vf } \big)   ~.
\label{Lag-tensor}
\eea
Here the chiral scalar $\vf$ and real linear $\cG$ superfields are related to  $\cG_{ij}$ as 
follows:
\be
\vf:= \cG_{\1\1}|~, \qquad \mathsf{G}:= 2i \,{\cG}_{\1\2}|~.
\ee
A short calculation gives
\be
\frac{\pa^2  L_{\rm IT}}{\pa \vf \pa\bar \vf} 
=-\frac{\pa^2  L_{\rm IT}}{\pa \mathsf{G}^2} = \frac{1}{\sqrt{\mathsf{G}^2 +4\vf \bar \vf } }
=  \frac{1}{2 \sqrt{\hf \cG^{ij}\cG_{ij} } }\Big|
~.
\ee
This result immediately allows us to construct a linearized action of the model,
$S^{(2)}$,
around a constant background $g^{ij}$,
\be
\cG^{ij} = g^{ij} + \qG^{ij}~, \qquad g^{ij} ={\rm const}~.
\ee
The linearized action is
\bea
S^{(2)} =   \frac{1}{2 g}  \int  {\rm d}^4 x \,{\rm d}^4\q \,\Big\{ 
\F \bar \F - \hf G^2\Big\}
\equiv  \frac{1}{2 g} S_{\rm T}
~, \qquad g:= \sqrt{\hf g^{ij}g_{ij} }~,
\eea
where the chiral scalar $\F$ and real linear $G$ superfields are defined by 
\be
\F= \bG_{\1\1}|~, \qquad {G}:= 2i \,{\bG}_{\1\2}|~.
\ee
Here $S_{\rm T}$ denotes the action for a massless $\cN=2$ tensor multiplet.

\section{Details of the $\mathcal N=1$ reduction}
We present here briefly the details for the $\N=1$ reduction. In the
first subsection, we give the explicit form of the $\N=1$ gauge transformations
in terms of the $\N=2$ gauge parameter. In the second, we give the form of the
$\N=1$ reduction for arbitrary values of the isospin vector $g_{ij}$.

\subsection{Derivation of the $\N=1$ gauge transformations}\label{N1gauge}
We relabel our two different Grassmann derivatives as
\begin{gather}
D_\alpha^{\ul 1} \rightarrow D_\alpha, \quad
D_\alpha^{\ul 2} \rightarrow \FF_\alpha
\end{gather}
and similarly for their conjugates.
Then the gauge transformation of $\qH$ is written
\begin{align}
\delta \qH = D^2 \qOmega_{\ul {11}} + 2 D^\alpha \FF_\alpha \qOmega_{\ul{12}} + \FF^2 \qOmega_{\ul{22}} 
+ \bar D^2 \bar\qOmega^{\ul{11}} + 2 \bar D_\dalpha \bFF^\dalpha \bar\qOmega^{\ul{12}}
     + \bFF^2 \bar\qOmega^{\ul{22}} ~.
\end{align}
${}$From this form it is clear that the higher $\theta_{\ul 2}$ components of
$\qOmega_{\ul{22}}$ and $\bar\qOmega^{\ul{22}}$ are available to eliminate the lower
$\theta_{\ul 2}$ components of $\qH$. In particular, it is possible to choose the Wess-Zumino
gauge \eqref{eq_WZgaugeH}. The residual components
of $\qH$ may be taken as in \eqref{eq_Hcomps}.

Because $\qOmega_{ij}$ has not been entirely fixed, the $\N=1$ superfields
must possess residual $\N=1$ gauge transformations. These may be derived by
direct computation, but we may first motivate their form
by considering the Noether coupling of $\qH$ to a conserved current $\mathcal J$.
Rewriting this coupling in $\N=1$ language yields
\begin{align}
\int \rd^4x \,\rd^8\theta \, \qH \mathcal J
= \int \rd^4x \,\rd^4\theta \left(
     H^{\dalpha \alpha} J_{\alpha \dalpha}
     + \Psi^\alpha J_\alpha + \bar\Psi_\dalpha J^\dalpha + 
     \hat U J
\right)~,
\end{align}
where the $\N=1$ currents are
\begin{subequations}
\begin{align}
J_{\alpha \dalpha} &= \frac{1}{4} [\FF_\alpha, \bFF_\dalpha] \mathcal J\vert
- \frac{1}{12} [D_\alpha,\bar D_\dalpha] \mathcal J\vert ~,\\
J_\alpha &= \FF_\alpha \mathcal J\vert ~, \\
J &= \mathcal J\vert~
\end{align}
\end{subequations}
and we have defined the combination
\begin{align}
\hat U := U + \frac{1}{12} [D^\alpha, \bar D^\dalpha] H_{\alpha \dalpha}~.
\end{align}

The $\N=2$ conservation condition is
\begin{align}
D^{ij} \mathcal J = \bar D_{ij} \mathcal J = 0~.
\end{align}
and implies for the $\N=1$ currents \cite{KT}
\begin{subequations}
\begin{gather}
D^\alpha J_{\alpha \dalpha} = \bar D^\dalpha J_{\alpha \dalpha} = 0 ~,\\
D^\alpha J_\alpha = \bar D^2 J_\alpha = \bar D_\dalpha J^\dalpha = D^2 J^\dalpha = 0~. \\
\bar D^2 J = D^2 J = 0~.
\end{gather}
\end{subequations}
These $\N=1$ conservation equations imply the corresponding gauge invariances \eqref{eq_N1gauge}.

It is a straightforward exercise to derive these gauge superfields in terms of
the $\mathcal N=2$ gauge parameter $\qOmega_{ij}$. First, the maintenance of Wess-Zumino
gauge fixes certain higher components of $\qOmega_{\ul{22}}$ and $\bar\qOmega^{\ul{22}}$:
\begin{subequations}\label{eq_WZ}
\begin{align}
\FF^2 \qOmega_{\ul{22}} + \bFF^2 \bar\qOmega^{\ul{22}}
&= - D^2 \qOmega_{\ul{11}} - \bar D^2 \bar\qOmega^{\ul{11}}
- 2 D^\alpha \FF_\alpha \qOmega_{\ul{12}}
- 2 \bar D_\dalpha \bFF^\dalpha \bar\qOmega^{\ul{12}}~, \\
\FF_\alpha \bFF^2 \bar\qOmega^{\ul{22}} &= -D^2 \FF_\alpha \qOmega_{\ul{11}} - \bar D^2 \FF_\alpha \bar\qOmega^{\ul{11}}
+ D_\alpha \FF^2 \qOmega_{\ul{12}} - 2 \FF_\alpha \bar D_\dbeta \bFF^\dbeta \bar\qOmega^{\ul{12}} ~,\\
\FF^2 \bFF^2 \bar\qOmega^{\ul{22}} &= -D^2 \FF^2 \qOmega_{\ul{11}} - \bar D^2 \FF^2 \bar\qOmega^{\ul{11}}
-2 \bar D_\dalpha \FF^2 \bFF^\dalpha \bar \qOmega^{\ul{12}}~.
\end{align}
\end{subequations}
where each equation should be understood as projected to $\theta_{\ul 2} = 0$.
These ensure that Wess-Zumino gauge is maintained by an otherwise arbitrary gauge transformation,
\begin{align}
\delta \qH\vert = \FF_\alpha \delta \qH \vert = \FF^2 \delta \qH \vert = 0~.
\end{align}

Calculating $\delta H_{\alpha \dalpha} = \frac{1}{4} [\FF_\alpha, \bFF_\dalpha] \delta \qH$
and imposing \eqref{eq_WZ}, one finds
\begin{align}
L_\alpha =&
- \frac{1}{2} \FF_\alpha \bFF^2 \bar\qOmega^{\ul{12}}
-\frac{1}{2} \bar D^\dbeta [\FF_\alpha, \bFF_\dbeta] \bar\qOmega^{\ul{11}}
+ \frac{1}{4} D_\alpha \bFF^2 \bar\qOmega^{\ul{22}}
+ \frac{1}{2} D_\alpha \bar D_\dbeta \bFF^\dbeta \bar\qOmega^{\ul{12}} \eol
& \quad
- \frac{1}{4} D_\alpha \FF^2 \qOmega_{\ul{22}}
+ \frac{1}{4} D^2 \FF_\alpha \qOmega_{\ul{12}} ~.
\end{align}
Similarly, one may calculate $\delta \Psi_\alpha$ and show that
\begin{align}
\Omega &=
- \frac{1}{8} \bFF^2 \FF^2 \qOmega_{\ul{12}}
-\frac{1}{4} D^\beta \bFF^2 \FF_\beta \qOmega_{\ul{11}}
+ \frac{1}{8} \bar D^2 \bFF^2 \bar\qOmega^{\ul{12}}~, \\
\Lambda_\alpha &=
\frac{1}{8} \bar D^2 \left(
\bFF^2 \FF_\alpha \bar\qOmega^{\ul{11}}
- D_\alpha \bFF^2 \bar\qOmega^{\ul{12}}
\right)~.
\end{align}

${}$For the $\N=1$ auxiliary superfield $U$, one finds
\begin{align}
\delta U = -\frac{i}{2} \partial^{\dalpha \alpha} \big(\bar D_\dalpha L_\alpha + D_\alpha L_\dalpha\big)
          + \Phi + \bar\Phi~,
\end{align}
where
\begin{align}
\Phi &= \frac{1}{16}\bar D^2 \left(
\FF^2 \bFF^2 \bar \qOmega^{\ul{11}} - 8 \Box \bar\qOmega^{\ul{11}}
+ 4i \partial^{\dalpha \alpha} D_\alpha \bFF_\dalpha \bar\qOmega^{\ul{12}}
+ \frac{1}{2} D^2 (\FF^2 \qOmega_{\ul{22}} - \bFF^2 \bar\qOmega^{\ul{22}})
\right) ~.
\end{align}
Using
\begin{align}
\delta \left([D_\beta, D_\dbeta] H^{\dbeta \beta}\right)
= 6i \partial^{\dalpha \alpha} \big(D_\alpha L_\dalpha + \bar D_\dalpha L_\alpha\big)
+ 2 D^2 (\bar D_\dalpha L^\dalpha)
+ 2 \bar D^2 (D^\alpha L_\alpha)
\end{align}
it follows that $\delta \hat U = \hat \Phi + \hat {\bar\Phi}$,
where
\begin{align}
\hat \Phi &= \Phi + \frac{1}{6} \bar D^2 D^\alpha L_\alpha ~.
\end{align}

\subsection{$\N=1$ reduction for arbitrary $g_{ij}$}\label{generalN1reduction}
To present the $\N=1$ action involving an arbitrary isospin unit vector $g_{ij}$,
we make the following identifications:
\begin{gather}
g_{\ul{11}} = x \gamma, \quad
g_{\ul{12}} = g_{\ul{21}} = i y, \quad
g_{\ul{22}} = x \bar \gamma \\
\gamma\bar\gamma = 1, \quad
x^2 + y^2 = 1
\end{gather}
where $x$ and $y$ are real parameters and $\gamma$ is a complex phase.
The constraint on $x$ and $y$ follows from the normalization condition
$g^{ij} g_{ij} = 2$.

We avoid giving the intermediate results for the actions $S_W$, $S_G$,
and $S_H$ as we do in the two special cases, but present merely the final
form of the action once all terms are collected together. We collect
first all terms quadratic in $\N=1$ components of $\qH$:
\begin{align}
S_{HH} &= \int \rd^4x\, \rd^4\theta \Bigg\{
     -\frac{1}{16} H^{\dalpha \alpha} D^\beta \bar D^2 D_\beta H_{\alpha \dalpha}
                   - \frac{1}{4} (\partial_{\alpha \dalpha} H^{\dalpha \alpha})^2  \eol
     &\quad
          + \frac{1}{64} (1+y^2) ([D_\alpha, \bar D_\dalpha] H^{\dalpha \alpha})^2
          -\frac{1}{8}(1+y^2) U [D_\beta, \bar D_\dbeta] H^{\dbeta \beta} \eol
     &\quad
     + \frac{i xy  \gamma}{8} D^\alpha \Psi_\alpha [D_\beta, \bar D_\dbeta] H^{\dbeta \beta}
     - \frac{i xy  \bar\gamma}{8} \bar D_\dalpha \bar \Psi^\dalpha [D_\beta, \bar D_\dbeta] H^{\dbeta \beta} \eol
     &\quad
     - \Psi^\alpha \bar D_\dalpha D_\alpha \bar\Psi^\dalpha
     - \frac{1}{4} w^2 \bar \Psi_\dalpha D^2 \bar\Psi^\dalpha
     - \frac{1}{4} \bar w^2 \Psi^\alpha \bar D^2 \Psi_\alpha \eol
     & \quad
     - \frac{1}{4} x^2 (\gamma D^\alpha \Psi_\alpha  - \bar\gamma \bar D_\dalpha \bar \Psi^\dalpha)^2
     + U \left(
     - \frac{i xy  \gamma}{2} D^\alpha \Psi_\alpha 
     + \frac{i xy  \bar\gamma}{2} \bar D_\dalpha \bar \Psi^\dalpha \right)\eol
     & \quad
     - \frac{1}{2} \left(1 + \frac{1}{2} x^2\right) U^2
     \Bigg\}
\end{align}
Note that when either $x$ or $y$ vanishes (i.e. the two cases we
have discussed in detail), the gravitino superfield $\Psi_\alpha$ decouples
from the $\N=1$ supergravity prepotential $H^{\dalpha \alpha}$. However, we see
that for the more general case, the action is more intricate in structure.

Next we present all terms involving both a supergravity field and a
compensator:
\begin{align}
S_{HC} &= \int \rd^4x\, \rd^4\theta \Bigg\{
     \frac{i x}{4} \,(\bar\gamma\eta - \gamma \bar \eta)\, \partial_{\alpha \dalpha} H^{\dalpha \alpha}
     + \frac{y}{8} \, G \, [D_\alpha, \bar D_\dalpha] H^{\dalpha \alpha} \eol
     &\quad
     + \frac{i}{2} (x \gamma G + y  \eta) D^\alpha \Psi_\alpha
     - \frac{i}{2} (x \bar\gamma G + y  \bar \eta) \bar D_\dalpha \bar \Psi^\dalpha
     - 2i \bar w \Psi^\alpha W_\alpha + 2i w \bar\Psi_\dalpha \bar W^\dalpha \eol
     &\quad
     + U \left(
          \frac{1}{2} x (\gamma \bar \eta + \bar\gamma \eta)
          - \frac{1}{2} y  G
          - \bar w \chi
          - w \bar\chi
     \right) \Bigg\}~.
\end{align}
We see that when both $x$ and $y$ are nonzero, the $\N=1$ supergravity
prepotential $H^{\dalpha \alpha}$ and the gravitino superfield $\Psi^\alpha$
share the same compensators.

The terms involving just the compensators are the same for all cases:
\begin{align}
S_{CC} &= - \int \rd^4x\, \rd^2\theta\, W^\alpha W_\alpha
     + \int \rd^4x\, \rd^4\theta\, \left(\frac{1}{4} G^2 - \frac{1}{2} \eta\bar\eta - \chi \bar \chi\right)~.
\end{align}

One may proceed as in the body of the paper to integrate out $U$;
however, because of the nontrivial couplings between the $\N=1$
supergravity prepotential $H_{\alpha \dalpha}$, the gravitino
superfield $\Psi_\alpha$, and their compensators, the generic action
does not take a particularly clean final form.

\section{The linearized hypermultiplet action}\label{omega_linearized}
The conventional formulation for a compensator hypermultiplet in harmonic
superspace \cite{GIKOS, GIOS} involves a doublet $q_a^+$ of analytic
superfields with an action
\begin{align}
S = \frac{1}{2} \int \rd\zeta^{(-4)} \, q^{a +} \CD^{++} q_a^+
\end{align}
where $\CD^{++}$ is the curved space generalization of $D^{++}$.
In terms of the supergravity prepotential $H$, $\CD^{++}$ may be
written
\begin{align}
\CD^{++} \Psi^{(n)} = D^{++} \Psi^{(n)} + (D^+)^4 (H D^{--} \Psi^{(n)})~, \qquad 
D^+_\ad \Psi^{(n)}
={\bar D}^+_\ad \Psi^{(n)}=0~,
\end{align}
where $\Psi^{(n)}$ is an arbitrary analytic superfield of $U(1)$ charge $n$.

This $q^+$ action can be rewritten involving a real hypermultiplet $\omega$
via the change of variables
\begin{align}
q_a^+ = u_a^+ \omega + u_a^- f^{++}
\end{align}
where $\omega$ and $f^{++}$ are analytic superfields.
In the resulting action, one may integrate out $f^{++}$ to end up with
\begin{align}
S = \int \rd\zeta^{(-4)} \bigg\{
     \frac{1}{2} \CD^{++} \omega \CD^{++} \omega
     + \frac{1}{2} \omega^2 \left(
          H^{(+4)}
          - \frac{1}{2} \CD^{++} \Gamma^{++}
          - \frac{1}{4} \Gamma^{++} \Gamma^{++}\right) \bigg\}
\end{align}
Here $H^{(+4)} = (D^+)^4 H$ and $\Gamma^{++} = (D^{+})^4(D^{--} H)$.

The supercurrent for this action in a Minkowski background
can be found by considering the terms first order in 
 the prepotential $H$.
One finds
\begin{align}
\delta S = \frac{1}{2} \int \rd^{4}x\, \rd^8\theta \,\rd u\, H \left(
     D^{++} \omega D^{--} \omega
     - \omega D^{--} D^{++} \omega
     + \omega^2
     \right)
\end{align}
which is valid for any gauge choice of $H$ and so the supercurrent is
\begin{align}
\mathcal J = \frac{1}{2} \left(
     D^{++} \omega D^{--} \omega
     - \omega D^{--} D^{++} \omega
     + \omega^2
     \right)
\end{align}
It is straightforward to check that this current obeys
both $(D^{+})^2 \mathcal J=0$ and $D^{++} \mathcal J = 0$ when the
$\omega$ hypermultiplet is on shell.

The full linearized action in this case is also quite easy to find since
$\omega$ is an unconstrained analytic superfield and so all $H$-dependence
appears explicitly in the covariant derivative and connection terms.
Linearizing $\omega$ about an on-shell background gives (in the central basis)
\begin{align}
\omega = \omega_0 + \omega^{ij} u_i^- u_j^+ + \qomega
\end{align}
where $\omega_0$ and $\omega^{ij}$ are constants and $\qomega$ is
unconstrained. It is convenient to write the second term in this expression
as $\omega^{-+}$.

The linearized action has the form
\begin{align}\label{eq_linomega}
S &= \frac{1}{2} \int \rd\zeta^{(-4)} (D^{++} \qomega)^2
     + \int \rd u\, \rd^4x \, \rd^8\theta\, \Bigg\{\qH (\omega_0 \qomega - 3 \omega^{-+} \qomega) \eol
     & \qquad
     + \frac{\omega_{ij} \omega_{kl}}{160} \qH D^{ij} \bar D^{kl} \qH 
     - \frac{\omega^{kl} \omega_{kl}}{480} \qH D^{ij} \bar D_{ij} \qH \Bigg\}
\end{align}
where we have chosen $\qH$ to be harmonic-independent. $\qomega$ varies
under the supergravity gauge transformation as an analytic density
of weight 1/2 \cite{KT}:
\begin{align}
\delta \qomega = (D^+)^4 D^{--} (l^{--} \omega)
     - \frac{1}{2} \omega (D^+)^4 D^{--} l^{--}
\end{align}
However, in order for the gauge choice for $\qH$ to be maintained, every
$\qOmega_{ij}$-transformation must be accompanied by a certain $l$-transformation
-- specifically,
\begin{align}
l^{--} = -(D^{-})^2 \qOmega^{-+} + 2 D^{\alpha -} D_{\alpha}^+ \qOmega^{--}
\end{align}
where $\qOmega^{-\pm} = \qOmega^{ij} u_i^- u_j^\pm$.
It follows that for this restricted class of gauge transformations (which is the
class that concerns us here)
\begin{align}\label{eq_domega}
\delta \qomega = (D^+)^4 (D^-)^2 \left(
     \frac{1}{2} \omega_0 \qOmega^{--}
     + \frac{1}{2} \omega^{-+} \qOmega^{--}
     - \omega^{--} \qOmega^{-+}
     \right)
\end{align}

While the generic background involves both an isosinglet $\omega_0$ and an isotriplet
$\omega^{ij}$, the theory dual to the improved tensor compensator should possess only an
isotriplet proportional to $g^{ij}$, and so we will consider the case where $\omega_0$ vanishes.

Now we may perform a duality transformation from the improved
tensor multiplet. Recall that we derived the form of the linearized
action for the improved tensor multiplet compensator only when coupled
to the Maxwell compensator. This meant that certain of the terms quadratic
in $\qH$ were necessarily of ambiguous origin: they could have arisen
from either the improved tensor or the Maxwell action. Indeed, the first
set of terms in \eqref{eq_SH} which were written with the coefficient $w\bar w$
could just have well been written with the coefficient $g$, since the action
was derived under the assumption that $g = w\bar w$. In order for the duality
transformation to reproduce \eqref{eq_linomega}, we will need to include
one such term $g \qH  D^{ij} \bar D_{ij} \qH$, with a constant coefficient $\lambda$ to be
determined:
\begin{multline}
S_{\rm IT} = -\frac{1}{4g} \int \rd\zeta^{(-4)} (\qG^{++})^2
     + \int \rd^4x\, \rd^8\theta\, \bigg\{\frac{1}{2g} g_{ij} \,\qG^{ij} \qH
     - \frac{1}{64g} g_{ij} g_{kl} \, \qH D^{ij} \bar D^{kl} \qH \\
     - \frac{\lambda g}{32} \qH D^{ij} \bar D_{ij} \qH\bigg\}
\end{multline}
We emphasize that this action is not by itself gauge-invariant, although it
does possess the property that its gauge variation is an $SU(2)$ invariant
(in the sense that it involves $g_{ij}$ only in the invariant combination
$g$) which can be cancelled by including the gauge variation of the
linearized Maxwell action under the assumption that $g = w\bar w$.

To perform the duality transformation, we rewrite the first two terms
in analytic superspace and introduce a Lagrange multiplier field $\qomega$
\begin{align}
\int \rd\zeta^{(-4)} \bigg\{-\frac{1}{4g} \qG^{++} \qG^{++}
     + \frac{3}{2g} g^{--} \,\qG^{++} (D^{+})^4 \qH
     + g^{-1/2} \qG^{++} D^{++} \qomega\bigg\}~,
\end{align}
which is an unconstrained analytic superfield enforcing
the constraint $D^{++} \qG^{++} = 0$. In order for this
action to be gauge invariant (up to terms independent of the
specific $SU(2)$ gauge choice of $g_{ij}$), $\qomega$ must
transform as
\begin{align}
\delta \qomega = g^{-1/2} (D^+)^4 (D^-)^2 \left(
     \frac{1}{2} \Omega^{--} g^{-+}
     - \Omega^{-+} g^{--} \right)
\end{align}
Comparing this result to \eqref{eq_domega}, we may tentatively
identify $g^{ij} / \sqrt{g}$ with $\omega^{ij}$ provided we work in the
case where $\omega_0 = 0$.

The dual action is found by integrating out $\qG^{++}$:
\begin{align}
\tilde S_{\rm IT} &=
     \int \rd\zeta^{-4} D^{++} \qomega D^{++} \qomega
     + \int \rd u\, \rd^4x\, \rd^8\theta \Bigg\{
     \frac{9}{4 g} (g^{--})^2 \qH (D^+)^4 \qH
     - \frac{6 g^{-+}}{\sqrt g} \qH \qomega \Bigg\} \eol
     & \qquad
     + \int \rd^4x\, \rd^8\theta \left(
     - \frac{1}{64g} g_{ij} g_{kl} \, \qH D^{ij} \bar D^{kl} \qH
     - \frac{\lambda g}{32} \qH D^{ij} \bar D_{ij} \qH\right)
\end{align}
The harmonic integral in the second term can be done, yielding
\begin{align}
\tilde S_{\rm IT} &=
     \int \rd\zeta^{-4} D^{++} \qomega D^{++} \qomega
     - \int \rd u\, \rd^4x\, \rd^8\theta \, \frac{6 g^{-+}}{\sqrt g} \qH \qOmega \eol
     & \qquad
     + \int \rd^4x\, \rd^8\theta \left\{
     \frac{1}{80 g} \qH D^{ij} \bar D^{kl} \qH  g_{ij} g_{kl}
     - \frac{g}{32} \left(\frac{3}{5} + \lambda\right) \qH D_{ij} \bar D^{ij} \qH \right\}
\end{align}
Comparing this to the linearized action for the $\omega$-hypermultiplet
in the case where $\omega_0$ vanishes \eqref{eq_linomega},
we find agreement up to an overall factor of 2
provided $\lambda = -1/3$ and $\omega^{ij} = g^{ij} / \sqrt g$.

In the supergravity formulation with the $\omega$ hypermultiplet, we have
several regimes to choose from for the background value. As we have just shown,
the case where $\omega_0$ vanishes is dual to a theory with an improved tensor
compensator. However, we may also choose the isotriplet $\omega_{ij}$ to vanish;
this would lead to a type-II supercurrent as discussed in the main body of the paper.

\footnotesize{

}


\begin{thebibliography}{66}


\bibitem{FZ}
 S.~Ferrara and B.~Zumino,
``Transformation properties of the supercurrent,''
Nucl.\ Phys.\  B {\bf 87}, 207 (1975).
  
\bibitem{OS}
V.~Ogievetsky and E.~Sokatchev,
``On vector superfield generated by supercurrent,''
Nucl.\ Phys.\  B {\bf 124}, 309 (1977).
   
\bibitem{FZ2}
S.~Ferrara and B.~Zumino,
``Structure of conformal supergravity,''  Nucl.\ Phys.\  B {\bf 134}, 301 (1978).

\bi{old}
J.~Wess and B.~Zumino,
``Superfield Lagrangian for supergravity,''
Phys.\ Lett.\  B {\bf 74}, 51 (1978);
K.~S.~Stelle and P.~C.~West,
``Minimal auxiliary fields for supergravity,''
Phys.\ Lett.\  B {\bf 74},  330 (1978);
S.~Ferrara and P.~van Nieuwenhuizen,
``The auxiliary fields of supergravity,''
Phys.\ Lett.\  B {\bf 74}, 333 (1978).

  
\bibitem{new}
V.~P.~Akulov, D.~V.~Volkov and V.~A.~Soroka,
``Generally covariant theories of gauge fields on superspace,'' 
Theor.\ Math.\ Phys.\  {\bf 31}, 285 (1977);
M.~F.~Sohnius and P.~C.~West,
``An alternative minimal off-shell version of N=1 supergravity,''
Phys.\ Lett.\  B {\bf 105}, 353 (1981).

\bi{non-min}
P.~Breitenlohner,
``Some invariant Lagrangians for local supersymmetry,''
Nucl.\ Phys.\ {\bf B124}, 500 (1977);
W.~Siegel and S.~J.~Gates Jr.
 ``Superfield supergravity,''  Nucl.\ Phys.\  B {\bf 147}, 77 (1979).
  
\bibitem{GGRS}
S.~J.~Gates Jr., M.~T.~Grisaru, M.~Ro\v{c}ek and W.~Siegel,
{\it Superspace, or One Thousand 
and One Lessons in Supersymmetry},
Benjamin/Cummings (Reading, MA),  1983, hep-th/0108200.

\bibitem{BK} I.~L.~Buchbinder and S.~M.~Kuzenko,
{\it Ideas and Methods of Supersymmetry and
Supergravity or a Walk Through Superspace},
IOP, Bristol, 1998.

\bibitem{Osborn}
  H.~Osborn,
  ``N = 1 superconformal symmetry in four-dimensional quantum field theory,''
  Annals Phys.\  {\bf 272}, 243 (1999)
  [arXiv:hep-th/9808041].

\bibitem{MSW}
 M.~Magro, I.~Sachs and S.~Wolf,
``Superfield Noether procedure,''  Annals Phys.\  {\bf 298}, 123 (2002)
 [arXiv:hep-th/0110131].


\bibitem{KS}
  Z.~Komargodski and N.~Seiberg,
``Comments on the Fayet-Iliopoulos term in field theory and supergravity,''
 JHEP {\bf 0906}, 007 (2009)  [arXiv:0904.1159 [hep-th]].

\bibitem{DT}
  K.~R.~Dienes and B.~Thomas,
  ``On the inconsistency of Fayet-Iliopoulos terms in supergravity theories,''
  Phys.\ Rev.\ D {\bf 81}, 065023 (2010). 


\bibitem{K-FI}
S.~M.~Kuzenko,
``The Fayet-Iliopoulos term and nonlinear self-duality,''
 Phys.\ Rev.\  D {\bf 81}, 085036 (2010)
 [arXiv:0911.5190 [hep-th]].
  
\bibitem{KS2}
  Z.~Komargodski and N.~Seiberg,
  ``Comments on supercurrent multiplets, supersymmetric field theories and
  supergravity,''
  JHEP {\bf 1007}, 017 (2010)
  [arXiv:1002.2228 [hep-th]].

\bibitem{K-var}
  S.~M.~Kuzenko,
  ``Variant supercurrent multiplets,''
  JHEP {\bf 1004}, 022 (2010)
  [arXiv:1002.4932 [hep-th]];

  
\bibitem{Butter}
  D.~Butter,
  ``Conserved supercurrents and Fayet-Iliopoulos terms in supergravity,''
  arXiv:1003.0249 [hep-th].

\bibitem{K-Noet} 
  S.~M.~Kuzenko,  ``Variant supercurrents and Noether procedure,''
  arXiv:1008.1877 [hep-th].
  
\bibitem{Sohnius}
  M.~F.~Sohnius,
  ``The multiplet of currents for N=2 extended supersymmetry,''
  Phys.\ Lett.\  B {\bf 81}, 8 (1979).
  
\bibitem{HST}
P.~S.~Howe, K.~S.~Stelle and P.~K.~Townsend,
``Supercurrents,''  Nucl.\ Phys.\  B {\bf 192}, 332 (1981).


\bibitem{Stelle}
K.~S.~Stelle,
``Extended supercurrents and the ultraviolet finiteness of N=4 supersymmetric
Yang-Mills theory,'' in {\it Quantum Structure of Space and Time}, M. J. Duff and C. J. Isham (Eds.), 
Cambridge University Press, Cambridge, 1982, p. 337.
 

\bibitem{GIKOS}
A.~Galperin, E.~Ivanov, S.~Kalitsyn, V.~Ogievetsky 
and E.~Sokatchev,
``Unconstrained N = 2 matter, Yang-Mills 
and supergravity theories 
in harmonic superspace,''
Class.\ Quant.\ Grav.\  {\bf 1}, 469 (1984).
 
 
\bibitem{Galperin:1987em}
A.~S.~Galperin, N.~A.~Ky and E.~Sokatchev,
``N=2 supergravity in superspace: Solution to the constraints,''
Class.\ Quant.\ Grav.\  {\bf 4}, 1235 (1987).

\bibitem{Galperin:1987ek}
A.~S.~Galperin, E.~A.~Ivanov, V.~I.~Ogievetsky and E.~Sokatchev,
``N=2 supergravity in superspace: Different versions and matter couplings,''
Class.\ Quant.\ Grav.\  {\bf 4}, 1255 (1987).

\bibitem{GIOS}
A.~S.~Galperin, E.~A.~Ivanov, V.~I.~Ogievetsky and E.~S.~Sokatchev,
{\it Harmonic Superspace}, Cambridge University Press, 
Cambridge, 2001.

\bibitem{KT}
S.~M.~Kuzenko and S.~Theisen,
 ``Correlation functions of conserved currents in N = 2 superconformal
theory,''  Class.\ Quant.\ Grav.\  {\bf 17}, 665 (2000)  [hep-th/9907107]. 

\bibitem{BS}
P.~Breitenlohner and M.~F.~Sohnius,
``Superfields, auxiliary fields, and tensor calculus for N=2 extended
supergravity,''
Nucl.\ Phys.\  B {\bf 165}, 483 (1980);
``An almost simple off-shell version of SU(2) Poincare supergravity,''
Nucl.\ Phys.\  B {\bf 178}, 151 (1981).

\bibitem{SSW}
  M.~F.~Sohnius, K.~S.~Stelle and P.~C.~West,
 ``Representations of extended supersymmetry,''
in {\it Superspace and Supergravity}, S. W. Hawking and M. Ro\v{c}ek (Eds.), 
Cambridge University Press, Cambridge, 1981, p. 283.  


\bibitem{GSW}
R.~Grimm, M.~Sohnius and J.~Wess,
``Extended supersymmetry and gauge theories,''
Nucl.\ Phys.\  B {\bf 133}, 275 (1978).

\bibitem{Mezincescu}
  L.~Mezincescu,
  ``On the superfield formulation of O(2) supersymmetry,''
  Dubna preprint JINR-P2-12572 (June, 1979).

\bibitem{rsg}
G.~Sierra and P.~K.~Townsend,
 ``An introduction to N=2 rigid supersymmetry,''
in {\it Supersymmetry and Supergravity 1983}, 
B. Milewski (Ed.), World Scientific, Singapore, 1983;
B.~de Wit, P.~G.~Lauwers, R.~Philippe, S.~Q.~Su and A.~Van Proeyen,
 ``Gauge and matter fields coupled to N = 2 supergravity,''
  Phys.\ Lett.\ B {\bf 134}, 37 (1984);
S.~J.~Gates Jr.,  ``Superspace formulation of new nonlinear sigma models,''
  Nucl.\ Phys.\ B {\bf 238}, 349 (1984).
  
  
\bibitem{deWvHVP}
 B.~de Wit, J.~W.~van Holten and A.~Van Proeyen,
 ``Transformation rules of N=2 supergravity multiplets,''
Nucl.\ Phys.\  B {\bf 167}, 186 (1980).

\bibitem{BdeRdeW}
  E.~Bergshoeff, M.~de Roo and B.~de Wit,
  ``Extended conformal supergravity,''
  Nucl.\ Phys.\  B {\bf 182}, 173 (1981).

\bibitem{deWPV}
B.~de Wit, R.~Philippe and A.~Van Proeyen,
``The improved tensor multiplet in N = 2 supergravity,''
Nucl.\ Phys.\ B {\bf 219}, 143 (1983).

\bibitem{LR}
U.~Lindstr\"om and M.~Ro\v{c}ek,
``Scalar tensor duality and N = 1, 2 nonlinear sigma models,''
Nucl.\ Phys.\  B {\bf 222}, 285 (1983).
  

\bibitem{deWR}
B.~de Wit and M.~Ro\v{c}ek,
``Improved tensor multiplets,''
Phys.\ Lett.\ B {\bf 109}, 439 (1982).

\bibitem{GKP}
  S.~J.~Gates Jr., S.~M.~Kuzenko and J.~Phillips,
  ``The off-shell (3/2,2) supermultiplets revisited,''
  Phys.\ Lett.\  B {\bf 576}, 97 (2003)
  [arXiv:hep-th/0306288].

\bibitem{RT}
  V.~O.~Rivelles and J.~G.~Taylor,
  ``Linearized N=2 superfield supergravity,''
  J.\ Phys.\ A  {\bf 15}, 163 (1982).

\bibitem{GS82}
S.~J.~Gates Jr. and W.~Siegel,
``Linearized N=2 superfield supergravity,''  Nucl.\ Phys.\  B {\bf 195}, 39 (1982).

\bibitem{GIO}
  A.~Galperin, E.~Ivanov and V.~Ogievetsky,
``Superspace actions and duality transformations for  N=2  tensor multiplets,''
  Phys.\ Scripta {\bf T15}, 176 (1987).


\bibitem{KLRT-M1}
S.~M.~Kuzenko, U.~Lindstr\"om, M.~Ro\v cek and G.~Tartaglino-Mazzucchelli,
``4D N=2 supergravity and projective superspace,'' 
JHEP {\bf 0809}, 051 (2008) [arXiv:0805.4683].

\bibitem{KLRT-M2}
 S.~M.~Kuzenko, U.~Lindstr\"om, M.~Ro\v{c}ek and G.~Tartaglino-Mazzucchelli,
``On conformal supergravity and projective superspace,''
JHEP {\bf 0908}, 023 (2009)
  [arXiv:0905.0063 [hep-th]].
  
\bibitem{K-dual08}
S.~M.~Kuzenko,
``On N = 2 supergravity and projective superspace: Dual formulations,''
Nucl.\ Phys.\  B {\bf 810}, 135 (2009)
[arXiv:0807.3381 [hep-th]].
    
\bibitem{Howe}
P.~S.~Howe,
``A superspace approach to extended conformal supergravity,''
Phys.\ Lett.\  B {\bf 100}, 389 (1981);
``Supergravity in superspace,''  Nucl.\ Phys.\  B {\bf 199}, 309 (1982).

\bibitem{Muller} 
M. M\"uller, {\it Consistent Classical Supergravity Theories},
(Lecture Notes in Physics, Vol. 336),
Springer, Berlin, 1989. 
  
\bibitem{KT-M}
S.~M.~Kuzenko and G.~Tartaglino-Mazzucchelli,
``Different representations for the action principle in 4D N = 2 supergravity,''
JHEP {\bf 0904}, 007 (2009)  [arXiv:0812.3464 [hep-th]].

\bibitem{Siegel-curved}
  W.~Siegel,
  ``Curved extended superspace from Yang-Mills theory a la strings,''
  Phys.\ Rev.\  D {\bf 53}, 3324 (1996)
  [hep-th/9510150].
  
\bibitem{AnB}
  I.~Antoniadis and M.~Buican,
  ``Goldstinos, supercurrents and metastable SUSY breaking in N=2
  supersymmetric gauge theories,''
  arXiv:1005.3012 [hep-th].
  
\bibitem{Z}
B.~M.~Zupnik,
``The action of the supersymmetric N = 2 gauge theory 
in harmonic superspace,''
Phys.\ Lett.\ B {\bf 183}, 175 (1987).

\bibitem{KLR}
A. Karlhede, U. Lindstr\"om and M. Ro\v cek,
``Self-interacting tensor multiplets in N = 2 superspace,''
Phys.\ Lett.\ B {\bf 147}, 297 (1984). 

\bibitem{LR-projective1}
U.~Lindstr\"om and M.~Ro\v{c}ek,
``New hyperk\"ahler  metrics  and new supermultiplets,''
 Commun.\ Math.\ Phys.\  {\bf 115}, 21 (1988).
 
\bibitem{LR-projective2}
U.~Lindstr\"om and M.~Ro\v{c}ek,
 ``N = 2 super Yang-Mills theory in projective superspace,''
Commun.\ Math.\ Phys.\  {\bf 128}, 191 (1990).

\bibitem{Siegel:1982wd}
  W.~Siegel,
   ``Chiral actions for N=2 supersymmetric tensor multiplets,''
  Phys.\ Lett.\  B {\bf 153}, 51 (1985).



\bibitem{FV}
  E.~S.~Fradkin and M.~A.~Vasiliev,
  ``Minimal set of auxiliary fields and S matrix for extended supergravity,''
  Lett.\ Nuovo Cim.\  {\bf 25}, 79 (1979);
  ``Minimal set of auxiliary fields in SO(2) extended supergravity,''
  Phys.\ Lett.\  B {\bf 85} (1979) 47.

\bibitem{dWvH}
  B.~de Wit and J.~W.~van Holten,
  ``Multiplets of linearized SO(2) supergravity,''
  Nucl.\ Phys.\  B {\bf 155}, 530 (1979).

\bibitem{GS80}
 S.~J.~Gates Jr. and W.~Siegel,
 ``(3/2, 1) superfield of O(2) supergravity,''
Nucl.\ Phys.\  B {\bf 164}, 484 (1980).

\bibitem{Ogievetsky:1976qb}
  V.~I.~Ogievetsky and E.~Sokatchev,
  ``On gauge spinor superfield,''
  JETP Lett.\  {\bf 23}, 58 (1976);


\bibitem{G-RRWLvU}
F.~Gonzalez-Rey, M.~Ro\v{c}ek, S.~Wiles, U.~Lindstr\"om and R.~von Unge,
``Feynman rules in N = 2 projective superspace. I: Massless  hypermultiplets,''
Nucl.\ Phys.\  B {\bf 516}, 426 (1998)
[arXiv:hep-th/9710250].


\bibitem{KT-M-ads}
  S.~M.~Kuzenko and G.~Tartaglino-Mazzucchelli,
  ``Field theory in 4D N=2 conformally flat superspace,''
  JHEP {\bf 0810}, 001 (2008)
  [arXiv:0807.3368 [hep-th]].

\bibitem{OS77}
  V.~I.~Ogievetsky and E.~Sokatchev,
 ``Superfield equations of motion,'' 
 J.\ Phys.\ A  {\bf 10}, 2021 (1977).

\bibitem{AB}
R.~Altendorfer and J.~Bagger,
 ``Dual supersymmetry algebras from partial supersymmetry breaking,''
Phys.\ Lett.\ B {\bf 460}, 127 (1999) 
[hep-th/9904213];
``Dual anti-de Sitter superalgebras from 
partial supersymmetry breaking,''
  Phys.\ Rev.\ D {\bf 61}, 104004 (2000) 
[hep-th/9908084].

\bibitem{BGLP}
I.~L.~Buchbinder, S.~J.~Gates, Jr., W.~D.~Linch and J.~Phillips,
``New 4D, N = 1 superfield theory: Model
of free massive superspin-3/2 multiplet,''
Phys.\ Lett.\ B {\bf 535}, 280 (2002) 
[hep-th/0201096];
``Dynamical superfield theory of free massive
superspin-1 multiplet,''
Phys.\ Lett.\ B {\bf 549}, 229 (2002) 
[hep-th/0207243].

\bibitem{Zinoviev}
  Y.~M.~Zinoviev,
 ``Massive spin-2 supermultiplets,''
hep-th/0206209.

\bibitem{GSS}
T.~Gregoire, M.~D.~Schwartz and Y.~Shadmi,
``Massive supergravity and deconstruction,''
JHEP {\bf 0407}, 029 (2004) 
[hep-th/0403224].

\bibitem{BGKP}
  I.~L.~Buchbinder, S.~J.~Gates, S.~M.~Kuzenko and J.~Phillips,
  ``Massive 4D, N = 1 superspin 1 and 3/2 multiplets and dualities,''
  JHEP {\bf 0502}, 056 (2005) 
[hep-th/0501199].

\bibitem{GKT-M}
  S.~J.~Gates Jr., S.~M.~Kuzenko and G.~Tartaglino-Mazzucchelli,
  ``New massive supergravity multiplets,''
  JHEP {\bf 0702}, 052 (2007)
  [arXiv:hep-th/0610333].




\end{thebibliography}
\end{document}